\documentclass[%aip,
 amsmath,amssymb,
preprint,%
nofootinbib
]{revtex4}

\usepackage{bm}
\usepackage{makecell}
\usepackage{multirow}
\usepackage{graphicx}
\usepackage{epstopdf}
\usepackage{tablefootnote}
\usepackage{subfigure}
\usepackage{amsmath}
\usepackage{caption}
\usepackage{color}

\usepackage{lscape}

\newcommand{\la}{\langle}
\newcommand{\ra}{\rangle}
\newcommand{\B}{{\cal B}}

\begin{document}

%\preprint{AIP/123-QED}

\title{Analysis of Hadronic Weak Decays of Charmed Baryons\\ in the Topological Diagrammatic Approach}
%[Sample title]{Sample Title:\\with Forced Linebreak\footnote{Error!}}% Force line breaks with \\
%\thanks{Footnote to title of article.}

\author{Huiling Zhong}
\author{Fanrong Xu}
\email{fanrongxu@jnu.edu.cn}
 \affiliation{Department of Physics, College of Physics $\&$ Optoelectronic Engineering, Jinan University, Guangzhou 510632, P.R. China}

\author{Hai-Yang Cheng}
\affiliation{Institute of Physics, Academia Sinica, Taipei, Taiwan 11529, Republic of China}

%\affiliation{ 
%Authors' institution and/or address%\\This line break forced with \textbackslash\textbackslash
%}%

\date{\today}% It is always \today, today,
             %  but any date may be explicitly specified
\small

\begin{abstract}
We perform a global fit to the experimental data of two-body charmed baryon decays based on the topological diagrammatic approach (TDA) and take
into account the phase shifts between $S$- and $P$-wave amplitudes as inspired by the recent BESIII measurement of the decay asymmetry in the decay $\Lambda_c^+\to \Xi^0K^+$. The TDA has the advantage that it is more intuitive, graphic and easier to implement model calculations. The measured branching fractions and decay asymmetries are well accommodated in the TDA except for a few modes, in particular, the predicted $\B(\Xi_c^0\to \Xi^-\pi^+)=(2.83\pm0.10)\%$ is larger than its current value.  The equivalence of the TDA and the irreducible SU(3) approach (IRA) is established.  We show that the number of the minimum set of tensor invariants in the IRA and the topological amplitudes in the TDA is the same and present their relations. 
The predicted magnitudes of $S$- and $P$-wave amplitudes and their phase shifts are presented for measured and yet-to-be-measured modes in both the TDA and IRA which can be tested in the near future.  Besides the decay $\Lambda_c^+\to \Xi^0 K^+$, there exist several modes which proceed only through  $W$-exchange. 
In particular, the observed channel $\Xi_c^0\to \Sigma^+ K^-$ 
should have phase shifts similar to that in $\Lambda_c^+\to \Xi^0 K^+$  and its decay asymmetry is predicted to be $-0.21\pm0.17$ which can be used to test our theoretical framework. In contrast, 
the TDA leads to a large $\alpha$ of order $-0.93$ for the decay $\Xi_c^+\to \Xi^0\pi^+$ even after the phase-shift effect is incorporated in the fit.

\end{abstract}

%Use showkeys class option if keyword
                              %display desired
\maketitle

\section{Introduction}
\label{sec:intro}
The progresses in the study of hadronic decays of charmed baryons, both experimentally and theoretically, had been very slow before 2014. Not only most of the experimental measurements were older ones (for a review, see Refs. \cite{Cheng:2009,Cheng:2015}), 
but also almost all the model calculations of charmed baryon decays were carried out before millennium. Indeed, theoretical interest in hadronic weak decays of charmed baryons peaked around the early 1990s and then faded away.

This situation was drastically changed after 2014 as there were several major breakthroughs in charmed-baryon experiments in regard to the weak decays of $\Lambda_c^+$ and $\Xi_c^{+,0}$ (for a review, see Refs. 
\cite{Cheng:2021qpd,Groote:2021pxt}). For example, the absolute branching fraction of $\Lambda_c^+\to pK^-\pi^+$, which is a benchmark for nearly all other branching fractions of $\Lambda_c^+$, has been measured by Belle \cite{Zupanc} and BESIII \cite{BES:pKpi} independently with much smaller uncertainties.
In 2015 BESIII has measured the absolute branching fractions of $\Lambda_c^+$ for more than a dozen of  decay modes directly for the first time \cite{BES:pKpi}.  This is a milestone in the study of hadronic charmed baryon 
decays. Likewise, Belle has reported the first measurements of the absolute branching fractions of $\Xi_c^0\to \Xi^-\pi^+$, $\Xi_c^+\to \Xi^-\pi^+\pi^+$ and
$\Xi_c^+\to pK^-\pi^+$ for the $\Xi_c^{+,0}$ systems \cite{Belle:Xic0,Belle:Xic+}. 

Considering the charmed baryon decay $\B_c\to\B_f+P$ with $P$ being a pseudoscalar meson and $\B_c$, $\B_f$ the charmed baryon and final-state baryon, respectively, its general decay amplitude reads 
\begin{eqnarray}
\label{eq:A&B}
M(\B_c\to \B_f+P)=i\bar u_f(A-B\gamma_5)u_c,
\end{eqnarray}
where $A$ and $B$ correspond to the parity-violating $S$-wave and parity-conserving $P$-wave amplitudes, respectively. In general, they receive both factorizable and nonfactorizable contributions 
\begin{equation} 
A=A^{\rm{fac}}+A^{\rm{nf}},\qquad
B=B^{\rm{fac}}+B^{\rm{nf}}.
\end{equation}
In the 1990s various approaches were developed to describe the nonfactorizable effects in hadronic decays of the charmed baryons $\Lambda_c^+$, $\Xi_c^{+,0}$ and $\Omega_c^0$. These include the covariant confined quark model, the pole model and current algebra (see Refs. \cite{Cheng:2021qpd,Groote:2021pxt} for references therein).

Besides the dynamical model calculations, a very promising approach is to use the approximate SU(3) flavor symmetry of QCD to describe the two-body nonleptonic decays of charmed baryons.  There exist two distinct ways in realizing the flavor symmetry, the irreducible SU(3) approach (IRA) and the topological diagram approach (TDA). They provide a powerful tool for a model-independent analysis. Among them, the IRA has become very popular in the past few years. In the IRA, SU(3) tensor invariants are constructed through the short-distance effective Hamiltonian, while in the TDA, the topological diagrams are classified according to the topologies in the flavor flow of weak decay diagrams with all strong-interaction effects included implicitly.  

Within the framework of the IRA, two-body nonleptonic decays of charmed baryons were first analyzed in Refs. \cite{Savage,Verma} followed by the   analysis of  Cabibbo-suppressed in Ref. \cite{Sharma}. After 2014, this approach became rather popular  \cite{Lu,Geng:Lambdac,Geng:2017mxn,Geng:2018,Hsiao:2019,He:2018joe,Jia:2019zxi}. 
However, the early studies of the IRA have overlooked the fact that charmed baryon decays are governed by several different partial-wave amplitudes which have distinct kinematic and dynamic effects. In other words, $S$- and $P$-waves were not distinguished in the early analysis and the IRA amplitudes are fitted only to the measured rates.
After the pioneer work in Ref. \cite{Geng:2019xbo}, it became a common practice to perform a global fit of both $S$- and $P$-wave parameters to the data of branching fractions and decay asymmetries \cite{Geng:2019awr,Geng:2020zgr,Huang:2021aqu,Zhong:2022exp,Xing:2023dni}.
Just like the case of hyperon decays, non-trival relative strong phases between $S$- and $P$-wave amplitudes may exist, but they were usually not considered in realistic model calculations of the decay asymmetry $\alpha$. 

The first analysis of two-body nonleptonic decays of antitriplet charmed baryons 
$\B_c(\bar 3)\to \B(8) M(8+1)$ within the framework of the TDA was performed by Kohara \cite{Kohara:1991ug}. A subsequent study was given by Chau, Cheng and Tseng (CCT) in Ref. \cite{Chau:1995gk}. Among the recent analyses in the TDA \cite{He:2018joe,Hsiao:2021nsc,Zhao:2018mov,Hsiao:2020iwc}, \footnote{The TDA analysis in Ref. \cite{Zhao:2018mov} did not assign the appropriate weight factors for the relevant topological diagrams in each decay.
}
there are 19 TDA amplitudes and 7 topological diagrams in Ref. \cite{He:2018joe}. Authors of Ref. \cite{Hsiao:2021nsc} followed the Kohara's scheme closely with 8 topological diagrams and 8 TDA amplitudes, but did not distinguish between  $S$- and $P$-wave contributions  and hence the topological amplitudes were fitted to the branching fractions only. Therefore, unlike the IRA, global fits to the rates and decay asymmetries are still absent in the TDA. 

Although the TDA has been applied very successfully to charmed meson decays \cite{CC,Cheng:2016,Cheng:2024hdo}, its application to charmed baryon decays is more complicated than the IRA. As stressed in Ref. \cite{He:2018joe}, it is easy to determine the independent amplitudes in the IRA, while the TDA gives some redundancy. Some of the amplitudes are not independent and therefore should be absorbed into other amplitudes. Nevertheless, the TDA has the advantage that it is more intuitive, graphic and easier to implement model calculations.  The extracted topological amplitudes by fitting to available data will enable us to probe the relative importance of different underlying decay mechanisms, and to relate one process to another at the topological amplitude level. 
In this work, we are going to show that the TDA is applicable to charmed baryon decays as well and it has the same number of independent amplitudes as that of the IRA.

The Cabibbo-favored mode $\Lambda_c^+\to\Xi^0 K^+$ which proceeds only through $W$-exchange deserves special attention. Early studies in 1990's indicated that its $S$- and $P$-wave amplitudes are very small due to strong cancellation between various terms (see e.g. Ref. \cite{Cheng:1993gf}). For example, the use of current algebra implies a vanishing $S$-wave in the SU(3) limit. Consequently, the calculated branching fraction is too small compared to experiment and the predicted $\alpha$ is zero owing to the vanishing $S$-wave amplitude. 
It is thus striking that the approach based on the IRA tends to predict a large decay asymmetry close to unity \cite{Geng:2019xbo,Zhong:2022exp,Xing:2023dni}. This long-standing puzzle was finally resolved by a recent BESIII measurement  \cite{BESIII:2023wrw}. Not only the decay asymmetry $\alpha_{\Xi^0K^+}=0.01\pm0.16$ was found to be consistent with zero, but also the measured Lee-Yang parameter $\beta_{\Xi^0K^+}=-0.64\pm0.69$ 
was nonzero,
implying a phase difference between $S$- and $P$-wave amplitudes, $\delta_P-\delta_S=-1.55\pm0.25$ or $1.59\pm0.25$ rad. 
Since $|\cos(\delta_P-\delta_S)|\sim 0.02$, this accounts for the smallness of $\alpha_{\Xi^0K^+}$. This first direct evidence supporting the existence of strong phases in the partial-wave amplitudes of hadronic charmed baryon decays plays a pivotal role in a further exploration of $C\!P$ violation in the charmed baryon sector.

Recently, a new analysis of charmed baryon decays based on the IRA that takes into account the phase shifts of the partial-wave amplitudes has been put forward in Ref. \cite{Geng:2023pkr}. 
In this work we shall perform a similar study within the framework of the TDA. 
Since the TDA has been applied very successfully to charmed meson decays, it is conceivable that the same approach is applicable to the charmed baryon sector.

The layout of this work is as follows. In Sec. II we first discuss the choice of octet-baryon wave wave functions. After writing down the general expression of the decay amplitudes in the TDA, we show that the number of independent amplitudes can be reduced through the K\"orner-Pati-Woo theorem and the removal of redundancy. The equivalence of the TDA and IRA is explicitly demonstrated in Sec. III. Sec. IV is devoted to the numerical analysis and fitting results. Sec. V comes to our conclusions. The relevant experimental results are collected in the Appendix. A short version of this work has been presented in Ref. \cite{Zhong:2024zme}.

\section{Formalism}
\label{sec:Forma}

Since baryons are made of three quarks in contrast to two quarks for the mesons, the application of TDA to the baryon case will inevitably lead to some complications. For example, the symmetry of the quarks in flavor space could be different. As stated in the Introduction, there exist two seemingly different analyses of two-body nonleptonic decays of antitriplet charmed baryons within the framework of the TDA: one by Kohara \cite{Kohara:1991ug} and the other by Chau, Cheng and Tseng (CCT) in Ref. \cite{Chau:1995gk}. The difference between Kohara and CCT lies in the choice of the wave functions of octet baryons:
\begin{eqnarray} \label{eq:wf8}
|{\cal B}^{m,k}(8)\rangle=a|\chi^m(1/2)_{A_{12}}\rangle|\psi^k(8)_{A_{12}}\rangle+ b|\chi^m(1/2)_{S_{12}}\rangle|\psi^k(8)_{S_{12}}\rangle
\end{eqnarray}
with $|a|^2+|b|^2=1$ in Ref. \cite{Chau:1995gk}, and
\begin{eqnarray} \label{eq:wf8tilde}
|\tilde{\cal B}^{m,k}(8)\rangle=\alpha|\chi^m(1/2)_{A_{12}}\rangle|\psi^k(8)_{A_{12}}\rangle+ \beta|\chi^m(1/2)_{A_{23}}\rangle|\psi^k(8)_{A_{23}}\rangle
\end{eqnarray}
in Ref. \cite{Kohara:1991ug}, where $\chi^m(1/2)_{A,S}$ are the spin parts of the wave function defined in Eq. (23) of Ref. \cite{Chau:1995gk} and
\begin{eqnarray}
|\psi^k(8)_{A_{12}}\rangle &=& \sum_{q_a,q_b,q_c}|[q_aq_b]q_c\rangle \langle[q_aq_b]q_c
|\psi^k(8)_{A_{12}}\rangle, \nonumber \\
|\psi^k(8)_{S_{12}}\rangle &=& \sum_{q_a,q_b,q_c}|\{q_aq_b\}q_c\rangle \langle\{q_aq_b\}q_c
|\psi^k(8)_{S_{12}}\rangle, 
\end{eqnarray}
are the octet baryon states that are antisymmetric and symmetric in the first two quarks, denoted by  $[~]$ and $\{\}$, respectively. As shown explicitly in Ref. \cite{Kohara:1997nu}, physics is independent of the convention one chooses. The TDA amplitudes expressed in the schemes with ${\cal B}^{m,k}(8)$ and $\tilde {\cal B}^{m,k}(8)$ are equivalent. Nevertheless, we prefer to use the bases $\psi^k(8)_{A_{12}}$ and $\psi^k(8)_{S_{12}}$ as they are orthogonal to each other, while $\psi^k(8)_{A_{12}}$ and $\psi^k(8)_{A_{23}}$ are not. 

To construct the decay amplitudes of ${\cal B}_c(\bar 3)\to {\cal B}(8)M(8+1)$ decays in the TDA, we first specify the building blocks.
The antitriplet charmed baryons are usually presented by
\begin{equation}
(\B_c)_{i}=\left(\Xi_c^0, -\Xi_c^+, \Lambda_c^+\right).
\end{equation}
For the purpose of constructing the TDA amplitudes, it is more convenient to
group them into an antisymmetric matrix $(\mathcal{B}_c)^{ij}=\epsilon^{ijk} (\mathcal{B}_c)_{k}$
\begin{equation}
(\mathcal{B}_c)^{ij}=\left(\begin{array}{ccc}
 0 & \Lambda_c^+ & \Xi_c^+ \\
-\Lambda_c^+& 0 & \Xi_c^0 \\
-\Xi_c^+& -\Xi_c^0 & 0 
\end{array}\right).
\end{equation}
The superscripts $i$ and $j$ also stand for the light quark flavors. For example,
$(\mathcal{B}_c)^{12}$ refers to the charmed baryon state $\Lambda_c^+$ with the quark content $cud$.  
The lowest-lying octet baryons $\B(8)$ are normally represented in the matrix form
\begin{equation}
(\mathcal{B}_8)^i_j=\left(\begin{array}{ccc}
\frac{1}{\sqrt{6}}\Lambda^0 + \frac{1}{\sqrt{2}}\Sigma^0 & \Sigma^+ & p \\
\Sigma^- & \frac{1}{\sqrt{6}}\Lambda^0 - \frac{1}{\sqrt{2}}\Sigma^0 & n \\
\Xi^-& \Xi^0& -\sqrt{\frac23}\Lambda^0 \end{array}\right). \\
\end{equation}
However, for the TDA purpose it is more convenient to introduce
the anti-symmetric tensor $\epsilon_{ijk}$ to write $(\mathcal{B}_8)_{i j k} = \epsilon_{ijl} (\mathcal{B}_8^{^T})^{l}_{k}$. The quark content of the baryon can be read from the subscript $ijk$. For example, $(\mathcal{B}_8)_{121} =(\mathcal{B}_8)_{u d u}=p$ and $(\mathcal{B}_8)_{122} =(\mathcal{B}_8)_{u d d}=n$.
Tensor form of the nonet pseudoscalar mesons $M(8+1)$ reads
\begin{equation}
\label{eq:meson1}
M^{i}_{j}=\left(\begin{array}{ccc}
\frac{\pi^0}{\sqrt{2}}+\frac{\eta_8}{\sqrt{6}}+\frac{\eta_1}{\sqrt{3}} & \pi^+ & K^+ \\
\pi^- & -\frac{\pi^0}{\sqrt{2}}+\frac{\eta_8}{\sqrt{6}}+\frac{\eta_1}{\sqrt{3}} & K^0 \\
K^- & \overline{K}^0 &  -\frac{2\eta_8}{\sqrt{6}}+\frac{\eta_1}{\sqrt{3}}
\end{array}\right),
\end{equation}
or
\begin{equation}
\label{eq:meson2}
M^{i}_{j}=\left(\begin{array}{ccc}
\frac{\pi^0+ \eta_{q}}{\sqrt{2}} & \pi^+ & K^+ \\
\pi^- & \frac{-\pi^0+ \eta_q}{\sqrt{2}} & K^0 \\
K^- & \overline{K}^0 & \eta_{s}
\end{array}\right),
\end{equation}
with 
\begin{equation}
\label{eq:eta81qs}
    \eta_{8}=\sqrt{\frac{1}{3}}\eta_{q}-\sqrt{\frac{2}{3}}\eta_{s},\quad
    \eta_{1}=\sqrt{\frac{2}{3}}\eta_{q}+\sqrt{\frac{1}{3}}\eta_{s}.
\end{equation}
The physical states $\eta$ and $\eta'$ are given by
\begin{equation}
\left(\begin{array}{c}
	\eta \\
	\eta'
\end{array}\right)=\left(\begin{array}{cc}
	\cos \phi & -\sin \phi\\
	\sin \phi & \cos \phi
\end{array}\right)\left(\begin{array}{l}
	\eta_{q} \\
	\eta_{s}
\end{array}\right)
=\left(\begin{array}{cc}
\cos \theta & -\sin \theta \\
\sin \theta & \cos \theta
\end{array}\right)\left(\begin{array}{c}
\eta_8 \\
\eta_1
\end{array}\right),
\end{equation}
where the mixing angles $\theta$ and $\phi$ are related through the relation $\theta=\phi-\arctan^{-1}\sqrt{2}$. For $\phi=40^\circ$, one will have $\theta=-15^\circ$. In Eqs. (\ref{eq:meson1}) and (\ref{eq:meson2}), the superscript $i$ stands for the quark flavor, while the subscript $j$ for the antiquark flavor.

\begin{figure}[t]
	\includegraphics[scale=1.00]{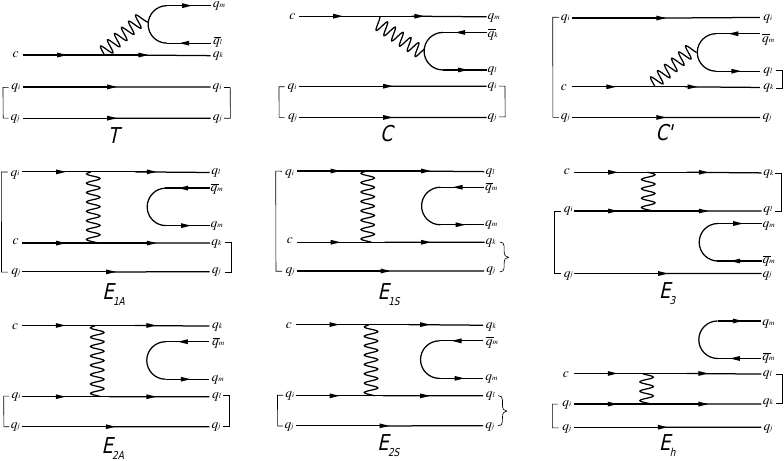}
	\caption{Topological diagrams contributing to ${\cal B}_c(\bar 3)\to {\cal B}(8)M(8+1)$ decays.}
	\label{Fig:TopDiag}
\end{figure}

In terms of the baryon and meson states, we follow Ref. \cite{He:2018joe} to write down the general expression of the decay amplitudes in the TDA:
\begin{equation}
\label{eq:TDAamp1}
\begin{aligned}
\mathcal{A}_{\rm TDA}=
& \quad T(\mathcal{B}_c)^{i j} H_l^{k m}M^l_m \left[
  b_{1}  \left(\mathcal{B}_8\right)_{ijk} 
+ b_{2} \left(\mathcal{B}_8\right)_{ikj} 
+ b_{3} \left(\mathcal{B}_8\right)_{jki} 
\right]\\
& +C(\mathcal{B}_c)^{i j} H_k^{m l}M^k_m\left[
  b_4 \left(\mathcal{B}_8\right)_{i j l} 
+ b_{5} \left(\mathcal{B}_8\right)_{i l j} 
+ b_{6} \left(\mathcal{B}_8\right)_{j l i} 
\right]\\
&+ C'(\mathcal{B}_c)^{i j} H_m^{k l}M^m_i\left[
b_{7} \left(\mathcal{B}_8\right)_{klj} 
+ b_{8} \left(\mathcal{B}_8\right)_{kjl} 
+ b_{9} \left(\mathcal{B}_8\right)_{ljk} 
\right]\\
&+E_{1}(\mathcal{B}_c)^{i j} H_i^{k l}M_l^m\left[
b_{10} \left(\mathcal{B}_8\right)_{jkm} 
+ b_{11}  \left(\mathcal{B}_8\right)_{jmk} 
+ b_{12}  \left(\mathcal{B}_8\right)_{kmj}
\right]\\
&+E_{2}(\mathcal{B}_c)^{i j} H_i^{k l}M_k^m\left[
b_{13} \left(\mathcal{B}_8\right)_{jlm} 
+ b_{14} \left(\mathcal{B}_8\right)_{jml} 
+ b_{15} \left(\mathcal{B}_8\right)_{lmj} 
\right]\\
&+E_{3}(\mathcal{B}_c)^{i j} H_i^{k l}M_j^m\left[
b_{16} \left(\mathcal{B}_8\right)_{klm} 
+ b_{17} \left(\mathcal{B}_8\right)_{kml} 
+ b_{18} \left(\mathcal{B}_8\right)_{lmk} 
\right] \\
&+E_{h}(\mathcal{B}_c)^{i j} H_i^{k l}M_m^m\left[
b_{19} \left(\mathcal{B}_8\right)_{jkl} 
+ b_{20} \left(\mathcal{B}_8\right)_{jlk} 
+ b_{21}  \left(\mathcal{B}_8\right)_{klj} 
\right],\\
\end{aligned}
\end{equation}
where the flavor indices of weak interactions are related to the $H$ matrix 
with the non-vanishing elements:
\begin{equation}
\label{eq:H}
H_{2}^{31}=V_{cs}^*V_{ud},\quad H_{3}^{31}=V_{cs}^*V_{us}, \quad H_{2}^{21}=V^*_{cd}V_{ud}, \quad H_{3}^{21}=V^*_{cd}V_{us}.
\end{equation}
The corresponding topological diagrams are depicted in Fig. \ref{Fig:TopDiag}: the external $W$-emission, $T$; the internal $W$-emission $C$; the inner $W$-emission $C'$; $W$-exchange diagrams $E_{1}$, $E_{2}$, $E_3$ and the hairpin diagram $E_h$. We shall see shortly that there exist two different types of $E_{1}$ and $E_{2}$ diagrams. Notice that $\mathcal{A}_{\rm T D A}$ in Eq. (\ref{eq:TDAamp1}) is the same as Eq. (80) of Ref. \cite{He:2018joe} except that we add one more term to $T$ and one to $C$ for the reason of completeness. 

At first sight, it appears that there are 21 amplitudes and 7 topological diagrams. However, because the two light quarks of the antitriplet charmed baryon are antisymmetric in flavor, so are the two spectator quarks $q_i$ and $q_j$  in diagrams $T$ and $C$ (see Fig. \ref{Fig:TopDiag}).   This implies that $b_3=-b_2$ and  $b_6=-b_5$.
Moreover, the final-state quarks $q_l$ and $q_k$  in topological diagrams $C'$, $E_3$ and $E_h$ must be antisymmetric in flavor owing to the K\"orner-Pati-Woo (KPW) theorem which states that the quark pair in a baryon produced by weak interactions is required to be antisymmetric in flavor in the SU(3) limit \cite{Korner:1970xq}. This amounts to having $b_9=-b_8$, $b_{18}=-b_{17}$ and $b_{20}=-b_{19}$. Furthermore, we notice that the combinations  of the coefficients $b_1+b_2$, $b_4+b_5$, $b_7+b_8$, $b_{16}+b_{17}$ and $b_{21}-b_{19}$ always appear in the decay amplitudes \cite{He:2018joe}. Hence, all of them can be absorbed into the topological amplitudes $T$, $C$, $C'$, $E_3$ and $E_h$, respectively. 
Consequently, the number of independent TDA amplitudes is reduced from 21 to 11.

In this work we shall employ the octet baryon wave function $\B^{m,k}(8)$ given by Eq. (\ref{eq:wf8}) and keep in mind that physics is independent of the choice of 
the baryon wave function, $\B^{m,k}(8)$ or $\tilde\B^{m,k}(8)$. Under this convention, we are forced to have $b_{12}=b_{11}$ and $b_{15}=b_{14}$. This means 
that the topological diagram $E_1$ is decomposed into two: $E_{1A}$ and $E_{1S}$
which are antisymmetric and symmetric in the quark pair $q_j$ and $q_k$, respectively. Likewise, the topological diagram $E_2$ is also decomposed into  $E_{2A}$ and $E_{2S}$ (see Fig. \ref{Fig:TopDiag}).
Absorbing the coefficients into the TDA amplitudes, the decay amplitudes of ${\cal B}_c(\bar 3)\to {\cal B}(8)M(8+1)$ in the TDA thus have the expressions:  
\begin{equation}
\label{Eq:TDAamp}
\begin{aligned}
 \mathcal{A}_{\rm T D A}%\\
=
& \quad T ({\mathcal{B}}_c)^{i j} H_l^{k m}\left(\mathcal{B}_8\right)_{i j k} M_m^l
+C (\mathcal{B}_c)^{i j} H_k^{m l}\left(\mathcal{B}_8\right)_{i j l} M_m^k
+ C' (\mathcal{B}_c)^{i j} H_m^{k l}\left(\mathcal{B}_8\right)_{klj} M_i^m \\
& +E_{1A} (\mathcal{B}_c)^{i j} H_i^{k l}\left(\mathcal{B}_8\right)_{jkm} M_l^m 
 + E_{1S} (\mathcal{B}_c)^{i j} H_i^{k l}M_l^m \left[\left(\mathcal{B}_8\right)_{jmk} 
+\left(\mathcal{B}_8\right)_{kmj} \right] \\
& +E_{2A} (\mathcal{B}_c)^{i j} H_i^{k l}\left(\mathcal{B}_8\right)_{jlm} M_k^m   + E_{2S} (\mathcal{B}_c)^{i j} H_i^{k l} M_k^m\left[\left(\mathcal{B}_8\right)_{jml}
+ \left(\mathcal{B}_8\right)_{lmj} \right] \\
&  +E_{3} (\mathcal{B}_c)^{i j} H_i^{k l}\left(\mathcal{B}_8\right)_{klm} M_j^m 
  +E_{h} (\mathcal{B}_c)^{i j} H_i^{k l}\left(\mathcal{B}_8\right)_{klj} M_m^m, \\
\end{aligned}
\end{equation}
where $E_{1A}=b_{10}E_1$ and $E_{1S}=b_{11}E_1$ and likewise  $E_{2A}=b_{13}E_2$ and $E_{2S}=b_{14}E_2$.  However, the final-state quarks $q_l$ and $q_k$ in the diagrams $E_{1A,1S}$ and $E_{2A,2S}$ are required by the KPW theorem be flavor antisymmetric in the SU(3) limit. Consequently, we are led to
\begin{equation}
    E_{2A}=-E_{1A}, \qquad E_{2S}=-E_{1S}.
\end{equation} 
As a result, the number of independent topological diagrams depicted in Fig. \ref{Fig:TopDiag} and the TDA amplitudes in Eq. (\ref{Eq:TDAamp}) is 7. 

\begin{table}[tp!]\footnotesize
\centering
\caption{TDA amplitudes for Cabibbo-favored (upper part) and singly Cabibbo-suppressed (lower part) ${\cal B}_c(\bar 3)\to {\cal B}(8)M(8+1)$ decays. Expressions of 
$\widetilde{\mathit{\rm TDA}}$ amplitudes are obtained using Eq. (\ref{eq:tildeTDA}).}
\label{tab:CFSCSamp}
%\resizebox{\textwidth}{!} 
\begin{tabular}{lll}
\hline
Channel & ~~~~~~TDA & ~~~~~$\widetilde{\mathit{\rm TDA}}$  \\
\hline
$\Lambda_c^+ \to \Lambda \pi^+$ &  
$\frac{1}{\sqrt{6}}(-4T+C'+E_{1A}+3E_{1S}-E_{3})$&
$\frac{1}{\sqrt{6}}(-4\Tilde{T}+\Tilde{C'}+\Tilde{E_{1}})$\\

$\Lambda_c^+ \to \Sigma^0 \pi^+$ &  
$\frac{1}{\sqrt{2}}(-C'-E_{1A}+E_{1S}+E_{3})$&
$\frac{1}{\sqrt{2}}(-\Tilde{C'}-\Tilde{E_{1}})$\\

$\Lambda_c^+ \to \Sigma^+ \pi^0$ &  
$\frac{1}{\sqrt{2}}(C'+E_{1A}-E_{1S}-E_{3})$&
$\frac{1}{\sqrt{2}}(\Tilde{C'}+\Tilde{E_{1}})$\\

%$\Lambda_c^+ \to \Sigma^+ \eta_{q}$ &  
%$\frac{1}{\sqrt{2}}(-C'+E_{1A}-E_{1S}-E_{3})$&-\\

%$\Lambda_c^+ \to \Sigma^+ \eta_{s}$ &  
%$2E_{2S}$& -\\

$\Lambda_c^+ \to \Sigma^+ \eta_{8}$ &  
$\frac{1}{\sqrt{6}}(-C'+E_{1A}+3E_{1S}-E_{3})$&
$\frac{1}{\sqrt{6}}(-\Tilde{C'}+\Tilde{E_{1}})$\\

$\Lambda_c^+ \to \Sigma^+ \eta_{1}$ &  
$\frac{1}{\sqrt{3}}(-C'+E_{1A}-3E_{1S}-E_{3}-3E_h)$&
$\frac{1}{\sqrt{3}}(-\Tilde{C'}+\Tilde{E_1}-3\Tilde{E_{h}})$\\

$\Lambda_c^+ \to \Xi^0 K^+$&
$E_{1A}+E_{1S}-E_{3}$&
$\Tilde{E_{1}}$\\

$\Lambda_c^+ \to p \bar{K}^0$&
$2C+2E_{1S}$&
$2\Tilde{C}$\\

$\Xi_c^0 \to \Lambda \bar{K}^0$&
$\frac{1}{\sqrt{6}}(2C-C'-E_{1A}+3E_{1S}+E_{3})$&
$\frac{1}{\sqrt{6}}(2\Tilde{C}-\Tilde{C'}-\Tilde{E_{1}})$\\

$\Xi_c^0 \to \Sigma^0 \bar{K}^0$&
$\frac{1}{\sqrt{2}}(2C+C'+E_{1A}+E_{1S}-E_{3})$&
$\frac{1}{\sqrt{2}}(2\Tilde{C}+\Tilde{C'}+\Tilde{E_{1}})$
\\

$\Xi_c^0 \to \Sigma^+ K^-$&
$-E_{1A}-E_{1S}+E_{3}$&
$-\Tilde{E_{1}}$\\

$\Xi_c^0 \to \Xi^0 \pi^0$&
$\frac{1}{\sqrt{2}}(-C'+2E_{1S})$&
$\frac{1}{\sqrt{2}}(-\Tilde{C'})$\\

%$\Xi_c^0 \to \Xi^0 \eta_q$&
%$\frac{1}{\sqrt{2}}(C'+2E_{1S})$& -\\

%$\Xi_c^0 \to \Xi^0 \eta_s$&
%$E_{2A}-E_{2S}+E_{3}$& -\\

$\Xi_c^0 \to \Xi^0 \eta_8$&
$\frac{1}{\sqrt{6}}(C'+2E_{1A}-2E_{3})$&
$\frac{1}{\sqrt{6}}(\Tilde{C'}+2\Tilde{E_{1}})$\\

$\Xi_c^0 \to \Xi^0 \eta_1$&
$\frac{1}{\sqrt{3}}(C'+3E_{1S}-E_{1A}+E_{3}+ 3E_h)$&
$\frac{1}{\sqrt{3}}(\Tilde{C'}-\Tilde{E_{1}}+3\tilde{E_h})$\\

$\Xi_c^0 \to \Xi^- \pi^+$&
$2T-2E_{1S}$&
$2\Tilde{T}$\\

$\Xi_c^+ \to \Sigma^+ \bar{K}^0$&
$-2C-C'$&
$-2\Tilde{C}-\Tilde{C'}$\\

$\Xi_c^+ \to \Xi^0 \pi^+$&
$-2T+C'$&
$-2\Tilde{T}+\Tilde{C'}$\\
\hline

$\Lambda_c^+ \to \Lambda K^+$ &  
$\frac{1}{\sqrt{6}}(-4T+C'-2E_{1A}+2E_{3})$&
$\frac{1}{\sqrt{6}}(-4\Tilde{T}+\Tilde{C'}-2\Tilde{E_{1}})$\\

$\Lambda_c^+ \to \Sigma^0 K^+$ & 
$\frac{1}{\sqrt{2}}(-C'+2E_{1S})$&
$\frac{1}{\sqrt{2}}(-\Tilde{C'})$\\

$\Lambda_c^+ \to \Sigma^+ K^0$ &
$-C'+2E_{1S}$&
$-\Tilde{C'}$\\

$\Lambda_c^+ \to p \pi^0$ &
$\frac{1}{\sqrt{2}}(2C+C'+E_{1A}+E_{1S}-E_{3})$&
$\frac{1}{\sqrt{2}}(2\Tilde{C}+\Tilde{C'}+\Tilde{E_{1}})$\\

% $\Lambda_c^+ \to p \eta_q$ &  
% $\frac{1}{\sqrt{2}}(-2C-C'+E_{1A}-E_{1S}+2E_{2S}-E_{3})$&
% -\\

% $\Lambda_c^+ \to p \eta_s$&
% $2C$&
% -\\

$\Lambda_c^+ \to p \eta_8$ &  
$\frac{1}{\sqrt{6}}(-6C+C'+E_{1A}-3E_{1S}-E_{3})$&
$\frac{1}{\sqrt{6}}(-6\Tilde{C}-\Tilde{C'}+\Tilde{E_{1}})$\\

$\Lambda_c^+ \to p \eta_1$&
$\frac{1}{\sqrt{3}}(-C'+E_{1A}-3E_{1S}-E_{3}-3E_{h})$&
$\frac{1}{\sqrt{3}}(-\Tilde{C'}+\Tilde{E_{1}}-3\Tilde{E_{h}})$\\

$\Lambda_c^+ \to n \pi^+$&
$-2T+C'+E_{1A}+E_{1S}-E_{3}$&
$-2\Tilde{T}+\Tilde{C'}+\Tilde{E_{1}}$\\

$\Xi_c^0 \to \Lambda \pi^0$&
$\frac{1}{2\sqrt{3}}(2C+2C'-3E_{1S}-E_{1A}+E_{3})$&
$\frac{1}{2\sqrt{3}}(2\Tilde{C}+2\Tilde{C'}-\Tilde{E_{1}})$\\

% $\Xi_c^0 \to \Lambda \eta_q$&
% $\frac{1}{2\sqrt{3}}(-2C-2C'-6E_{1S}-E_{2A}+3E_{2S}-E_{3})$&
% -\\

% $\Xi_c^0 \to \Lambda \eta_s$&
% $\frac{1}{\sqrt{6}}(2C-C'-2E_{2A}-2E_{3})$&
% -\\

$\Xi_c^0 \to \Lambda \eta_8$&
$\frac{1}{2}(-2C-3E_{1S}-E_{1A}+E_{3})$&
$\frac{1}{2}(-2\Tilde{C}-\Tilde{E_{1}})$\\

$\Xi_c^0 \to \Lambda \eta_1$&
$\frac{1}{\sqrt{2}}(-C'-3E_{1S}+E_{1A}-E_{3}-3E_{h})$&
$\frac{1}{\sqrt{2}}(-\Tilde{C'}+\Tilde{E_{1}}-3\Tilde{E_{h}})$\\

$\Xi_c^0 \to \Sigma^0 \pi^0$&
$\frac{1}{2}(2C+3E_{1S}+E_{1A}-E_{3})$&
$\frac{1}{2}(2\Tilde{C}+\Tilde{E_{1}})$\\

% $\Xi_c^0 \to \Sigma^0 \eta_q$&
% $\frac{1}{2}(-2C+2E_{1S}+E_{2A}+E_{2S}+E_{3})$&
% -\\

% $\Xi_c^0 \to \Sigma^0 \eta_s$&
% $\frac{1}{\sqrt{2}}(2C+C'-2E_{2S})$&
% -\\

$\Xi_c^0 \to \Sigma^0 \eta_8$&
$\frac{1}{2\sqrt{3}}(-6C-2C'-3E_{1S}-E_{1A}+E_{3})$&
$\frac{1}{2\sqrt{3}}(-6\Tilde{C}-2\Tilde{C'}-\Tilde{E_{1}})$\\

$\Xi_c^0 \to \Sigma^0 \eta_1$&
$\frac{1}{\sqrt{6}}(C'+3E_{1S}-E_{1A}+E_{3}+3E_{h})$&
$\frac{1}{\sqrt{6}}(\Tilde{C'}-\Tilde{E_1}+3\Tilde{E_{h}})$\\

$\Xi_c^0 \to \Sigma^+ \pi^-$&
$E_{1A}+E_{1S}-E_{3}$&
$\Tilde{E_1}$\\

$\Xi_c^0 \to \Sigma^- \pi^+$&
$-2T+2E_{1S}$&
$-2\Tilde{T}$\\

$\Xi_c^0 \to \Xi^0 K^0$&
$C'+E_{1A}-E_{1S}-E_{3}$&
$\Tilde{C'}+\Tilde{E_1}$\\

$\Xi_c^0 \to \Xi^- K^+$&
$2T-2E_{1S}$&
$2\Tilde{T}$\\

$\Xi_c^0 \to p K^-$&
$-E_{1A}-E_{1S}+E_{3}$&
$-\Tilde{E_1}$\\

$\Xi_c^0 \to n \bar{K}^0$&
$-C'-E_{1A}+E_{1S}+E_{3}$&
$-\Tilde{C'}-\Tilde{E_1}$\\

$\Xi_c^+ \to \Lambda \pi^+$&
$\frac{1}{\sqrt{6}}(2T-2C'+E_{1A}+3E_{1S}-E_{3})$&
$\frac{1}{\sqrt{6}}(2\Tilde{T}-2\Tilde{C'}+\Tilde{E_{1}})$\\

$\Xi_c^+ \to \Sigma^0 \pi^+$&
$\frac{1}{\sqrt{2}}(-2T-E_{1A}+E_{1S}+E_{3})$&
$\frac{1}{\sqrt{2}}(-2\Tilde{T}-\Tilde{E_{1}})$\\

$\Xi_c^+ \to \Sigma^+ \pi^0$&
$\frac{1}{\sqrt{2}}(-2C+E_{1A}-E_{1S}-E_{3})$&
$\frac{1}{\sqrt{2}}(-2\Tilde{C}+\Tilde{E_{1}})$\\

% $\Xi_c^+ \to \Sigma^+ \eta_q$&
% $\frac{1}{\sqrt{2}}(2C+E_{1A}-E_{1S}-E_{3})$&
% -\\

% $\Xi_c^+ \to \Sigma^+ \eta_s$&
% $-2C-C'+2E_{2S}$&
% -\\

$\Xi_c^+ \to \Sigma^+ \eta_8$&
$\frac{1}{\sqrt{6}}(6C+2C'+E_{1A}+3E_{1S}-E_{3})$&
$\frac{1}{\sqrt{6}}(6\Tilde{C}+2\Tilde{C'}+\Tilde{E_{1}})$\\

$\Xi_c^+ \to \Sigma^+ \eta_1$&
$\frac{1}{\sqrt{3}}(-C'+E_{1A}-3E_{1S}-E_{3}-3E_{h})$&
$\frac{1}{\sqrt{3}}(-\Tilde{C'}+\Tilde{E_{1}}-3\Tilde{E_{h}})$\\

$\Xi_c^+ \to \Xi^0 K^+$&
$-2T+C'+E_{1A}+E_{1S}-E_{3}$&
$-2\Tilde{T}+\Tilde{C'}+\Tilde{E_{1}}$\\

$\Xi_c^+ \to p \bar{K}^0$&
$-C'+2E_{1S}$&
$-\Tilde{C'}$\\
\hline
\end{tabular}
\end{table}

If we choose $\tilde{B}^{m,k}$ given in Eq. (\ref{eq:wf8tilde}) as the octet baryon wave function, we will have $b_{11}=b_{14}=0$. Consequently, the topological diagrams $E_{1S}$ and $E_{2S}$ in Fig. \ref{Fig:TopDiag} are replaced by $E'_{1A}$ and $E'_{2A}$, respectively. They are antisymmetric in the quark pairs
$(q_k, q_m)$ and $(q_l, q_m)$, respectively. In this case, the TDA amplitudes read
\begin{equation}
\begin{aligned}
 \tilde{\mathcal{A}}_{\rm T D A}%\\
=
& \quad T ({\mathcal{B}}_c)^{i j} H_l^{k m}\left(\mathcal{B}_8\right)_{i j k} M_m^l
+C (\mathcal{B}_c)^{i j} H_k^{m l}\left(\mathcal{B}_8\right)_{i j l} M_m^k\\
& + C' (\mathcal{B}_c)^{i j} H_m^{k l}\left(\mathcal{B}_8\right)_{klj} M_i^m
 +E_{1A} (\mathcal{B}_c)^{i j} H_i^{k l}\left(\mathcal{B}_8\right)_{jkm} M_l^m \\
& + E'_{1A} (\mathcal{B}_c)^{i j} H_i^{k l}\left(\mathcal{B}_8\right)_{kmj} M_l^m
 +E_{2A} (\mathcal{B}_c)^{i j} H_i^{k l}\left(\mathcal{B}_8\right)_{jlm} M_k^m  \\
& + E'_{2A} (\mathcal{B}_c)^{i j} H_i^{k l}\left(\mathcal{B}_8\right)_{lmj} M_k^m 
  +E_{3} (\mathcal{B}_c)^{i j} H_i^{k l}\left(\mathcal{B}_8\right)_{klm} M_j^m \\
 & +E_{h} (\mathcal{B}_c)^{i j} H_i^{k l}\left(\mathcal{B}_8\right)_{klj} M_m^m.
\end{aligned}
\end{equation}
Just like the previous case,
the final-state quarks $q_l$ and $q_k$ in the $W$-exchange diagrams $E_{1A,2A}$ and $E'_{1A,2A}$ are required be flavor antisymmetric. This implies that
\begin{equation}
\label{eq:EinIRA}
    E_{2A}=-E_{1A}, \qquad E'_{2A}=-E'_{1A}.
\end{equation} 
Note that $\tilde{\mathcal{A}}_{\rm T D A}$ is the starting point of Ref. \cite{Hsiao:2021nsc} for analyzing charmed baryon decays in the TDA. However, the relations shown in Eq. (\ref{eq:EinIRA}) were not utilized by the authors.  

\begin{table}\footnotesize
\centering
\caption{Same as Table \ref{tab:CFSCSamp} except for doubly Cabibbo-suppressed charmed baryon weak decays.}
\label{tab:DCSamp}
\begin{tabular}{lll}
\hline
Channel & TDA & $\widetilde{\mathit{\rm TDA}}$\\
\hline
$\Lambda_c^+ \to p K^0$&
$2C+C'$&
$2\Tilde{C}+\Tilde{C'}$\\

$\Lambda_c^+ \to n K^+$&
$2T-C'$&
$2\Tilde{T}-\Tilde{C'}$\\

$\Xi_c^0 \to \Lambda K^0$&
$\frac{1}{\sqrt{6}}(2C+2C'+2E_{1A}-2E_{3})$&
$\frac{1}{\sqrt{6}}(2\Tilde{C}+2\Tilde{C'}+2\Tilde{E_{1}})$\\

$\Xi_c^0 \to \Sigma^0 K^0$&
$\frac{1}{\sqrt{2}}(2C+2E_{1S})$&
$\frac{1}{\sqrt{2}}(2\Tilde{C})$\\

$\Xi_c^0 \to \Sigma^- K^+$&
$2T-2E_{1S}$&
$2\Tilde{T}$\\

$\Xi_c^0 \to p \pi^-$&
$-E_{1A}-E_{1S}+E_{3}$&
$-\Tilde{E_{1}}$\\

$\Xi_c^0 \to n \pi^0$&
$\frac{1}{\sqrt{2}}(E_{1S}+E_{1A}-E_{3})$&
$\frac{1}{\sqrt{2}}(\Tilde{E_{1}})$\\

%$\Xi_c^0 \to n \eta_q$&
%$\frac{1}{\sqrt{2}}(2E_{1S}+E_{2A}-E_{2S}+E_{3})$&-\\

%$\Xi_c^0 \to n \eta_s$&
%$C'$& -\\

$\Xi_c^0 \to n \eta_8$&
$\frac{1}{\sqrt{6}}(-2C'+3E_{1S}-E_{1A}+E_{3})$&
$\frac{1}{\sqrt{6}}(-2\Tilde{C'}-\Tilde{E_{1}})$\\

$\Xi_c^0 \to n \eta_1$&
$\frac{1}{\sqrt{3}}(C'+3E_{1S}-E_{1A}+E_{3}+3E_h)$&
$\frac{1}{\sqrt{3}}(\Tilde{C'}-\Tilde{E_{1}}+3\Tilde{E}_{h})$\\

$\Xi_c^+ \to \Lambda K^+$&
$\frac{1}{\sqrt{6}}(-2T+2C'+2E_{1A}-2E_{3})$&
$\frac{1}{\sqrt{6}}(-2\Tilde{T}+2\Tilde{C'}+2\Tilde{E_{1}})$\\

$\Xi_c^+ \to \Sigma^0 K^+$&
$\frac{1}{\sqrt{2}}(2T-2E_{1S})$&
$\frac{1}{\sqrt{2}}(2\Tilde{T})$\\

$\Xi_c^+ \to \Sigma^+ K^0$&
$-2C-2E_{1S}$&
$-2\Tilde{C}$\\

$\Xi_c^+ \to p \pi^0$&
$\frac{1}{\sqrt{2}}(-E_{1A}-E_{1S}+E_{3})$&
$\frac{1}{\sqrt{2}}(-\Tilde{E_{1}})$\\

%$\Xi_c^+ \to p \eta_q$&
%$\frac{1}{\sqrt{2}}(-E_{1A}+E_{1S}-2E_{2S}+E_{3})$& -\\

%$\Xi_c^+ \to p \eta_s$&
%$C'$& -\\

$\Xi_c^+ \to p \eta_8$&
$\frac{1}{\sqrt{6}}(-2C'-E_{1A}+3E_{1S}+E_{3})$&
$\frac{1}{\sqrt{6}}(-2\Tilde{C'}-\Tilde{E_{1}})$\\

$\Xi_c^+ \to p \eta_1$&
$\frac{1}{\sqrt{3}}(C'-E_{1A}+3E_{1S}+E_{3}+3E_h)$&
$\frac{1}{\sqrt{3}}(\Tilde{C'}-\Tilde{E_{1}}+3\tilde{E}_{h})$\\

$\Xi_c^+ \to n \pi^+$&
$-E_{1A}-E_{1S}+E_{3}$&
$-\Tilde{E_{1}}$\\
\hline
\end{tabular}
% }
\label{tab:DCSdecay}
\end{table}

Working out Eq. (\ref{Eq:TDAamp}) for ${\cal B}_c(\bar 3)\to {\cal B}(8)M(8+1)$ decays,  we obtain the TDA decay amplitudes listed in Tables \ref{tab:CFSCSamp}
and \ref{tab:DCSamp}. It is straightforward to show that
the expressions of the TDA amplitudes agree with CCT \cite{Chau:1995gk} through the following relations:
\begin{eqnarray}
&& {\cal A}_A=-4T, \quad {\cal B}_A=2C', \quad {\cal B}'_A=-4C, \quad {\cal C}_{1A}=2E_3,   \nonumber \\
&& {\cal C}_{2A}=-{\cal C}'_{A}=2E_{1A}, \quad 
{\cal C}_{2S}=-{\cal C}'_S=-2\sqrt{3}E_{1S},
\end{eqnarray}
where ${\cal A}_A$, ${\cal B}_A$, ${\cal B}'_A$, ${\cal C}_{1A}$, ${\cal C}_{2A}$, ${\cal C}'_{A}$, 
${\cal C}_{2S}$ and ${\cal C}'_S$ are the TDA amplitudes defined in Ref. \cite{Chau:1995gk}. The agreement is non-trivial in view of the different methodologies adopted in \cite{Chau:1995gk} and here. 
Among the 7 TDA amplitudes given in Eq. (\ref{Eq:TDAamp}), there still exist 2 redundant degrees of freedom through the redefinitions \cite{Chau:1995gk}:
\begin{eqnarray}
\label{eq:tildeTDA}
&& \tilde T=T-E_{1S}, \quad \tilde C=C+E_{1S},\quad \tilde C'=C'-2E_{1S}, \nonumber\\   
%&& \tilde E_2=E_{2A}+E_{2S}+E_3, \quad \tilde E_3=E_{1S}+E_{2S},   \nonumber \\
&& \tilde E_1=E_{1A}+E_{1S}-E_3, \quad \tilde E_h=E_h+2E_{1S}. 
\end{eqnarray}
A closer look of the TDA amplitudes of Cabibbo-favored, singly-Cabibbo-suppressed and doubly-Cabibbo-supprerssed decays given in Tables \ref{tab:CFSCSamp} and \ref{tab:DCSamp} 
shows that $E_{1S}$ can be absorbed by $T$, $C$,  $C'$, $E_1$ and $E_h$, as shown in the above equation. Hence, the redundant $E_{1S}$  can be eliminated. Also the amplitude $E_3$ is always accompanied by $E_{1A}+E_{1S}$. Consequently, it can be absorbed by the combination of $E_{1A}+E_{1S}$.  As a result, among the seven topological amplitudes $T$, $C$, $C'$,  $E_{1A}$,  $E_h$, $E_{1S}$ and $E_3$, the last two are redundant degrees of freedom and can be omitted through redefinitions.

In Tables \ref{tab:CFSCSamp}
and \ref{tab:DCSamp} we shall use $\widetilde{\mathit{\rm TDA}}$ to denote
the tilde TDA amplitudes. By now it is clear that the minimum set of the topological amplitudes is 5. This is in agreement with the number of tensor invariants found in the IRA \cite{Geng:2023pkr} (see also Sec. III). 

For the $\eta$ and $\eta'$ final states, we have exhibited the TDA amplitudes in Tables \ref{tab:CFSCSamp} and \ref{tab:DCSamp} in the $\eta_8$ and $\eta_1$ basis. Of course, one can also work in the $\eta_q$ and $\eta_s$ basis. For example,
\begin{equation}
\begin{aligned}
& {\cal A}(\Lambda_c^+\to\Sigma^+\eta_q)=\frac{1}{\sqrt{2}}(-C'+E_{1A}-E_{1S}-E_3-2E_h)=\frac{1}{\sqrt{2}}(-\tilde{C}'+\tilde{E}_1-2\tilde{E}_h), \\
& {\cal A}(\Lambda_c^+\to\Sigma^+\eta_s)=-2E_{1S}-E_h=-\tilde{E}_h.
\end{aligned}
\end{equation}
Applying Eq. (\ref{eq:eta81qs}), we obtain
\begin{equation}
\begin{aligned}
& {\cal A}(\Lambda_c^+\to\Sigma^+\eta_8)=\frac{1}{\sqrt{6}}(-\tilde{C}'+\tilde{E}_1), \\
& {\cal A}(\Lambda_c^+\to\Sigma^+\eta_1)=\frac{1}{\sqrt{3}}(-\tilde{C}'+\tilde{E}_1-3\tilde{E}_h).
\end{aligned}
\end{equation}
We see that the hairpin diagram $E_h$ \cite{Li,hairpin,Chau:1990ht} contributes to the SU(3)-singlet $\eta_1$ but not to $\eta_8$ as it should be.

Many sum-rule relations can be derived from Table \ref{tab:CFSCSamp}, for example, 
\begin{eqnarray}
\label{eq:sumrule}
{\tau_{\Lambda_c^+}\over\tau_{\Xi_c^0}}\B(\Xi_c^0\to\Xi^-\pi^+) &=& 3\B(\Lambda_c^+\to \Lambda\pi^+)+\B(\Lambda_c^+\to \Sigma^0\pi^+)-{1\over \sin^2\theta_C}\B(\Lambda_c^+\to n\pi^+),  \nonumber \\
{\tau_{\Xi_c^0}\over\tau_{\Lambda_c^+}}\B(\Lambda_c^+\to p \overline{K}^0) &=& 3\B(\Xi_c^0\to \Lambda\overline{K}^0)+\B(\Xi_c^0\to \Sigma^0\overline{K}^0)-{1\over \sin^2\theta_C}\B(\Xi_c^0\to \Xi^0 K^0).
\end{eqnarray}
The first relation was first derived in Ref. \cite{Geng:2023pkr}. It is very useful to constrain the branching fraction of $\Xi_c^0\to\Xi^-\pi^+$.

\section{Equivalence of TDA and IRA}
In this section we will consider the general SU(3)-invariant decay amplitudes in the IRA. To demonstrate the equivalence between the TDA and IRA, we need to show that the number of the minimum set of tensor invariants in the IRA and the topological amplitudes in the TDA is the same. If the number of minimum independent amplitudes is different, 
the global fit to the data will yield different fitting results for branching fractions, decay asymmetries and phase shifts. 

Consider the effective Hamiltonian responsible for the $\Delta C=1$ weak transition
\begin{equation}
\begin{aligned}
{\cal H}_{\rm eff}= & {G_F\over \sqrt{2}}\sum_{q_1,q_2}^{d,s} V_{cq_1}V_{uq_2}(c_1 O_1^{q_1q_2}+c_2 O_2^{q_1q_2})+ h.c.\\
= & {G_F\over \sqrt{2}}\sum_{q_1,q_2}^{d,s} V_{cq_1}V_{uq_2}(c_+ O_+^{q_1q_2}+c_- O_-^{q_1q_2})+ h.c.,
\end{aligned}
\end{equation}
where $O_1^{q_1q_2}=(\bar uq_2)(\bar q_1 c)$, $O_2^{q_1q_2}=(\bar q_1q_2)(\bar u c)$ with $(\bar q_1q_2)\equiv \bar q_1\gamma_\mu(1-\gamma_5)q_2$, $O_\pm\equiv {1\over 2}(O_1\pm O_2)$ and $c_\pm\equiv c_1\pm c_2$. 
Under flavor SU(3) symmetry, the operators $O_-$ and $O_+$ transform as the irreducible representations of {\bf 6} and ${\bf \overline{15}}$, respectively. 
Notice that $O_+$ ($O_-$) is symmetric (antisymmetric) in the color indices of light quarks.  Since the Wilson coefficients $c_-\gg c_+$, it has been assumed
the sextet ${\bf 6}$ dominance over $\overline{\bf 15}$ in the literature. Under this hypothesis, one would have the relation \cite{Lu}
\begin{eqnarray} \label{eq:npi}
\B(\Lambda_c^+\to n\pi^+)=2\B(\Lambda_c^+\to p\pi^0),
\end{eqnarray}
and the sum rule
\begin{eqnarray} \label{eq:SR}
\B(\Lambda_c^+\to p\overline{K}^0) &=& 3\B(\Lambda_c^+\to \Lambda\pi^+)+\B(\Lambda_c^+\to \Sigma^0\pi^+)-{1\over \sin^2\theta_C}\B(\Lambda_c^+\to n\pi^+).
\end{eqnarray}
This sum rule is identical to the first one in Eq. (\ref{eq:sumrule}) except for the left-hand-side term. However, both the relation (\ref{eq:npi}) and the sum rule (\ref{eq:SR}) are not borne out by experiment. This indicates that the contributions from ${\cal H}_{\rm eff}(\overline{15})$ cannot be neglected.

We follow Ref. \cite{He:2018joe} to write down the general SU(3) invariant decay amplitudes in the IRA: \footnote{The relations between the IRA amplitudes in Ref. \cite{He:2018joe} and ours are: $a_{1}=A_6^T$, $a_{2}=B_6^T$, $a_{3}=C_6^T$, $a_{4}=E_6^T$, $a_{5}=D_6^T$, 
$a_{6}=A_{15}^T$, $a_{7}=B_{15}^T$, $a_{8}=C_{15}^T$, $a_{9}=E_{15}^T$ and $a_{10}=D_{15}^T$. }
\begin{equation}
\begin{aligned}
\label{eqs:IRAamp}
\mathcal{A}_{\rm I R Aa}= & \quad a_{1}\left({\B}_c\right)_i\left(H_6\right)_j^{i k}\left({\B}_8\right)_k^j M_l^l
+a_{2}\left({\B}_c\right)_i\left(H_6\right)_j^{i k}\left({\B}_8\right)_k^l M_l^j
+a_{3}\left({\B}_c\right)_i\left(H_6\right)_j^{i k}\left({\B}_8\right)_l^j M_k^l \\
& +a_{4}\left({\B}_c\right)_i\left(H_6\right)_l^{j k}\left({\B}_8\right)_j^i M_k^l
+a_{5}\left({\B}_c\right)_i\left(H_6\right)_l^{j k}\left({\B}_8\right)_j^l M_k^i\\
&+a_{6}\left({\B}_c\right)_i\left(H_{\overline{15}}\right)_j^{i k}\left({\B}_8\right)_k^j M_l^l
+a_{7}\left({\B}_c\right)_i\left(H_{\overline{15}}\right)_j^{i k}\left({\B}_8\right)_k^l M_l^j
+a_{8}\left({\B}_c\right)_i\left(H_{\overline{15}}\right)_j^{i k}\left({\B}_8\right)_l^j M_k^l\\
&+a_{9}\left({\B}_c\right)_i\left(H_{\overline{15}}\right)_l^{j k}\left({\B}_8\right)_j^i M_k^l
+a_{10}\left({\B}_c\right)_i\left(H_{\overline{15}}\right)_l^{j k}\left({\B}_8\right)_j^l M_k^i.\\
\end{aligned}
\end{equation}
For the explicit expressions of $\left(H_{6}\right)_k^{i j}$ and $\left(H_{\overline{15}}\right)_k^{i j}$, see Ref. \cite{He:2018joe}. 
The first five terms associated with $H_6$ are not totally independent as one of them is redundant through the redefinition. It should be stressed that the redefinition is not unique. For example, we will consider the following redefinitions 
\begin{equation}
\label{eq:redef2}
 a_1'=a_1-a_5, \quad   a_2'=a_2+a_5, \quad a_3'=a_3+a_5, \quad a_4'=a_4+a_5,  
\end{equation} 
which amounts to deleting the $a_5$ term. There is another set of redefinitions
adopted in Ref. \cite{He:2018joe}
\begin{equation}
\label{eq:redef1}
 a_1''=a_1+a_2, \quad   a_2''=a_2-a_3, \quad a_3''=a_3-a_4, \quad a_5''=a_5+a_3.  
\end{equation}
It is straightforward to check that the IRA amplitudes given in Tables 14-16 of Ref. \cite{He:2018joe} can be expressed in terms of $a_1', a_2',a_3',a_4'$ or
$a_1'', a_2'',a_3'',a_5''$ together with $a_6,\cdots,a_{10}$.

As for the five terms associated with $H_{\overline{15}}$ in Eq. (\ref{eqs:IRAamp}), four of them are prohibited by the KPW theorem, namely, $a_6=a_7=a_8=a_{10}=0$ \cite{Geng:2018rse,Geng:2023pkr}. To see this, we follow the argument presented in Ref. \cite{Geng:2018rse} closely. Consider the decay amplitude of $\B_c\to \B_8 M$ induced by $H_{\overline{15}}$ or the operator $O_+$. For the nonfactorizable contributions of $O_+$, the relevant matrix elements are $\la \B^*|O_+|\B_c\ra$ and 
$\la \B_8|O_+|\B^*\ra$ after considering the pole contributions from the intermediate baryon pole $\B^*$. Since $O_+$ is symmetric in color indices while baryons are antisymmetric, we are led to $\la\B_i|O_+|\B_j\ra=0$, which is one of the ingredients for the derivation of the KPW theorem. 
Hence, we are left with the factorizable contributions of $\la \B_8 M|O_+|\B_c\ra$, which can be inferred from the factorizable terms $T$ and $C$ given in Eq. (\ref{Eq:TDAamp}), 
\begin{equation}
\la \B_8 M|{\cal H}_{\rm eff}|\B_c\ra^{\rm fact}=T ({\mathcal{B}}_c)^{i j} H_l^{k m}\left(\mathcal{B}_8\right)_{i j k} M_m^l
+C (\mathcal{B}_c)^{i j} H_k^{m l}\left(\mathcal{B}_8\right)_{i j l} M_m^k
\end{equation}
The $H$ matrix given in Eq. (\ref{eq:H}) is related to $H_{6}$ and $H_{\overline{15}}$ via
\begin{equation}
\label{eq:H15&6}
H_k^{ij}={1\over 2}\left[H(\overline{15})_k^{ij}+{1\over 2}\epsilon^{ijl}H(6)_{kl}\right],
\end{equation}
where the expression of $H(6)_{kl}$ can be found in e.g. Ref. \cite{Geng:2018rse}.
Projecting out the factorizable contributions from $H(\overline{15})$, we obtain
\begin{equation}
\la \B_8 M|{\cal H}_{\rm eff}(\overline{15})|\B_c\ra^{\rm fact}=(T+C) \left({\B}_c\right)_i\left(H_{\overline{15}}\right)_l^{j k}\left({\B}_8\right)_j^i M_k^l.
\end{equation}
Comparing this with Eq. (\ref{eqs:IRAamp}) leads to
\begin{equation}
a_9=T+C=\tilde{T}+\tilde{C}, \qquad a_6=a_7=a_8=a_{10}=0. 
\end{equation}

In the IRA we thus have five independent SU(3) tensor invariants $a_1', a_2', a_3', a_4'$ and $a_9'=a_9$, in agreement with Ref. \cite{Geng:2023pkr}. In the TDA we find the minimum set of the topological amplitudes: $\tilde{T}, \tilde{C}, \tilde{C'}, \tilde{E_1}$
and $\tilde{E_h}$. By comparing the $\widetilde{\mathit{\rm TDA}}$ amplitudes in Tables \ref{tab:CFSCSamp} and \ref{tab:DCSamp} with the IRA amplitudes given in Tables 14-16 of Ref. \cite{He:2018joe}, 
we arrive at the relations
\begin{equation}
\label{eqs:IRAtoTDAtilde}
\begin{aligned}
&a_{1}-a_{5}=-\Tilde{E_{h}}, \qquad a_{2}+a_{5}=-\Tilde{C'},\\
&a_{3}+a_{5}=\Tilde{E_{1}}, \qquad a_{4}+a_{5}=\Tilde{T}-\Tilde{C}-\Tilde{C'},\\
&a_{9}=\Tilde{T}+\Tilde{C}, \qquad a_{6}=a_{7}=a_{8}=a_{10}=0,\\
\end{aligned}
\end{equation}
and hence
\begin{equation}
\label{eqs:TDAtildetoIRA}
\begin{aligned}
&\Tilde{T}=\frac{1}{2}(-a_{2}+a_{4}+a_{9}),\qquad \Tilde{C}=\frac{1}{2}(a_{2}-a_{4}+a_{9}),\\
&\Tilde{C'}=-a_{2}-a_{5}, \qquad \Tilde{E_{1}}=a_{3}+a_{5}, \qquad \Tilde{E_{h}}=-a_{1}+a_{5}.
\end{aligned}
\end{equation}
Therefore, we have the correspondence
\begin{equation}
\begin{aligned}
\label{eq:tildeTDAIRAprime}
& \Tilde{T}=\frac{1}{2}(-a_{2}'+a_{4}'+a_{9}'),\qquad \Tilde{C}=\frac{1}{2}(a_2'-a_{4}'+a_{9}'),\\
& \Tilde{C'}=-a_{2}', \qquad \Tilde{E_{1}}=a_{3}', \qquad \Tilde{E_{h}}=-a_{1}'.
\end{aligned}
\end{equation}
in terms of the redefinitions given in Eq. (\ref{eq:redef2}). The equivalence between the TDA and IRA is thus established. 

There is another set of the IRA amplitudes given in Ref. \cite{Geng:2023pkr}
\begin{equation}
\label{eq:IRAb}
\begin{aligned}
\mathcal{A}_{\rm I R A b}= & \quad \tilde{f}^a\left({\B}_c\right)^{ik}\left(H_6\right)_{ij}\left({\B}_8\right)_k^j M_l^l+\tilde{f}^b \left({\B}_c\right)^{ik}\left(H_6\right)_{ij}\left({\B}_8\right)_k^l M_l^j+\tilde{f}^c \left({\B}_c\right)^{ik}\left(H_6\right)_{ij}\left({\B}_8\right)_l^j M_k^l \\
& +\tilde{f}^d \left({\B}_c\right)^{kl}\left(H_6\right)_{ij}\left({\B}_8\right)_k^i M_l^j+\tilde{f}^e\left({\B}_c\right)_j\left(H_{\overline{15}}\right)_l^{i k}\left({\B}_8\right)_i^j M_k^l.
\end{aligned}
\end{equation}
The equivalence between $\widetilde{\rm TDA}$, IRAa and IRAb leads to the relations:
\begin{equation}
\begin{aligned}
\label{eq:tildeTDAIRA}
&\Tilde{T}
=\frac{1}{2}( \tilde f^b+\tilde f^e),\qquad
\Tilde{C}
=\frac{1}{2}(-\tilde f^b+\tilde f^e), \\
& \Tilde{C'}=\tilde f^b-\tilde f^d,\qquad
\Tilde{E_{1}}=-\tilde f^c,\qquad
\Tilde{E_{h}}=\tilde f^a,\\
\end{aligned}
\end{equation}
and
\begin{equation}
\begin{aligned}
\label{eq:tildeIRAprime}
a_1'=\tilde f^a,  \qquad a_2'=-\tilde f^b+\tilde f^d,\qquad
a_3'=-\tilde f^c, \qquad
a_4'=\tilde f^d, \qquad
a_9'=\tilde f^e. 
\end{aligned}
\end{equation}

It should be stressed that since we will only fit to the observed branching fractions and decay asymmetries, the amplitudes are subject to sign ambiguity. Hence,  the relations (\ref{eq:tildeTDAIRAprime}), (\ref{eq:tildeTDAIRA}) and (\ref{eq:tildeIRAprime}) also hold if all the relations are flipped in sign simultaneously. For example,
\begin{equation}
\begin{aligned}
&\Tilde{T}
=-\frac{1}{2}( \tilde f^b+\tilde f^e),\qquad
\Tilde{C}
=\frac{1}{2}(\tilde f^b-\tilde f^e), \\
& \Tilde{C'}=-\tilde f^b+\tilde f^d,\qquad
\Tilde{E_{1}}=\tilde f^c,\qquad
\Tilde{E_{h}}=-\tilde f^a\\
\end{aligned}
\end{equation}
are also valid.

\section{Numerical Analysis and Results}
\label{sec:Num}
As there are 5 independent tilde TDA amplitudes given in Eq. (\ref{eq:tildeTDA}),  we have totally 19 unknown parameters to describe the magnitudes and the phases of the respective $S$- and $P$-wave amplitudes, namely, 
\begin{eqnarray}
\label{eq:19parameters}
&& |\tilde T|_Se^{i\delta_S^{\tilde T}}, \quad |\tilde C|_Se^{i\delta_S^{\tilde C}}, \quad |\tilde C'|_Se^{i\delta_S^{\tilde C'}},  \quad |\tilde E_{1}|_Se^{i\delta_S^{\tilde E_{1}}}, \quad |\tilde E_h|_Se^{i\delta_S^{\tilde E_h}},\nonumber \\
&& |\tilde T|_Pe^{i\delta_P^{\tilde T}}, \quad |\tilde C|_Pe^{i\delta_P^{\tilde C}}, \quad |\tilde C'|_Pe^{i\delta_P^{\tilde C'}}, \quad |\tilde E_{1}|_Pe^{i\delta_P^{\tilde E_{1}}}, \quad |\tilde E_h|_Pe^{i\delta_P^{\tilde E_h}},
\end{eqnarray}
collectively denoted by $|X_i|_Se^{i\delta^{X_i}_S}$ and $|X_i|_Pe^{i\delta^{X_i}_P}$,
where the subscripts $S$ and $P$ denote the $S$- and $P$-wave components of each TDA amplitude. 
Since there is an overall phase which can be omitted, we shall set $\delta_S^{\tilde T}=0$. Hence, we are left with 19 parameters. Likewise, for the tidle IRA amplitudes given in Eq. (\ref{eq:IRAb}), we also have
\begin{eqnarray}
\label{eq:IRA19parameters}
&& |\tilde f^a|_Se^{i\delta_S^{\tilde f^a}}, \quad |\tilde f^b|_Se^{i\delta_S^{\tilde f^b}}, \quad |\tilde f^c|_Se^{i\delta_S^{\tilde f^c}},  \quad |\tilde f^d|_Se^{i\delta_S^{\tilde f^d}}, \quad |\tilde f^e|_Se^{i\delta_S^{\tilde f^e}},\nonumber \\
&& |\tilde f^a|_Pe^{i\delta_P^{\tilde f^a}}, \quad |\tilde f^b|_Pe^{i\delta_P^{\tilde f^b}}, \quad |\tilde f^c|_Pe^{i\delta_P^{\tilde f^c}},  \quad |\tilde f^d|_Pe^{i\delta_P^{\tilde f^d}}, \quad |\tilde f^e|_Pe^{i\delta_P^{\tilde f^e}}.
\end{eqnarray}
We shall aslo set $\delta_S^{\tilde f^a}=0$. Of course, physics is independent of which phase is removed. Indeed, one can check that the phase differences $\delta_P^{X_i}-\delta_S^{X_i}$ or $\delta_P^{\tilde f^x}-\delta_S^{\tilde f^x}$
remain the same no matter which phase is set to zero.
Notice that the number of available experimental observables has increased to 30 by the end of 2023.
To pursue a set of proper parameters, 
the $\chi^2$ function in the following maximum likelihood analysis is defined as
\begin{equation}
\chi^2=\left[\mathcal{O}_{\text{theor}}(c_i)-\mathcal{O}_{\text{expt}}\right]^{\text{T}}\Sigma^{-1}
\left[\mathcal{O}_{\text{theor}}(c_i)-\mathcal{O}_{\text{expt}}\right],
\end{equation}
in which $c_i$ are the fitted $19$ input parameters, $\mathcal{O}_{\text{theor}, \text{expt}}$
stand for the $30$ theoretical and experimental observables.
The  $30$-dimensional general error matrix $\Sigma$ can be taken diagonal by
neglecting correlations among different observables and only 
incorporating pure experimental errors here.

\begin{table}[t]
\caption{ Fitted tilde TDA (upper) and IRA (lower) amplitudes collectively denoted by $X_i$. We have set $\delta^{\tilde{T}}_S=0$
and $\delta^{\tilde{f^a}}_S=0$.
} 
\label{tab:coeff}
\vspace{-0.1cm}
\begin{center}
\renewcommand\arraystretch{1}
% \resizebox{\textwidth}{!} 
% {
\begin{tabular}%{\textwidth}
{c| c r r r }
\hline
&$|X_i|_S$&$|X_i|_P$~~~ &$\delta^{X_i}_{S}$~~~~~ &$\delta^{X_i}_{P}$~~~~ \\
&\multicolumn{2}{c}{$(10^{-2}G_{F}~{\rm GeV}^2)$}&\multicolumn{2}{c}{$(\text{in radian})$} \\
\hline
$\Tilde{T}$&
~$2.37\pm0.41$&$16.56\pm0.69$&
-- ~~~~~~~&$2.76\pm0.32$\\
$\Tilde{C}$&
~$1.04\pm1.08$&$13.82\pm0.58$&
$-1.97\pm0.79$&$-0.37\pm0.44$\\
$\Tilde{C'}$&
~$2.59\pm0.95$&$24.97\pm1.67$&
$0.29\pm0.19$&$2.86\pm0.36$\\
$\Tilde{E_{1}}$&
~$4.10\pm0.20$&$2.56\pm2.21$&
$1.18\pm0.38$&$-0.96\pm0.43$\\
$\Tilde{E_{h}}$&
~$1.54\pm1.22$&$19.16\pm3.00$&
$-1.35\pm0.60$&$0.37\pm0.41$\\
\hline
$\tilde{f}^a$&
~$0.81\pm1.89$&$23.02\pm4.04$&
-- ~~~~~~~&$2.12\pm1.03$\\
$\tilde{f}^b$&
~$2.89\pm1.50$&$30.56\pm1.30$&
$2.03\pm0.61$&$-1.78\pm0.98$\\
$\tilde{f}^c$&
~$4.20\pm0.18$&$1.95\pm2.21$&
$-0.06\pm1.03$&$-2.68\pm1.16$\\
$\tilde{f}^d$&
~$0.98\pm0.90$&$7.25\pm2.07$&
$2.72\pm1.29$&$-2.55\pm1.00$\\
$\tilde{f}^e$&
~$2.06\pm0.62$&$4.73\pm2.11$&
$1.09\pm0.99$&$-0.94\pm0.99$\\
\hline
\end{tabular}
% }
\end{center}
\end{table}

In terms of the $S$- and $P$-wave amplitudes given in Eq. (\ref{eq:A&B}) 
and their phases $\delta_S$ and $\delta_P$, respectively, the decay rate and Lee-Yang parameters read
\begin{equation}
\begin{split}
&\Gamma = \frac{p_c}{8\pi}\frac{(m_i+m_f)^2-m_P^2}{m_i^2}\left(|A|^2
+ \kappa^2|B|^2\right),\\
& \alpha=\frac{2\kappa |A^*B|\cos(\delta_P-\delta_S)}{|A|^2+\kappa^2 |B|^2},~~
\beta=\frac{2\kappa |A^*B|\sin(\delta_P-\delta_S)}{|A|^2+\kappa^2 |B|^2},~~
\gamma=\frac{|A|^2-\kappa^2 |B|^2}{|A|^2+\kappa^2 |B|^2},
\end{split}
\label{eq:kin}
\end{equation}
where  $p_c$ is the c.m. three-momentum in the rest
frame of initial baryon and the auxiliary parameter $\kappa$ is 
defined as $\kappa=p_c/(E_f+m_f)=\sqrt{(E_f-m_f)/(E_f+m_f)}$. 
The phase shift between $S$- and $P$-wave amplitudes can be deduced explicitly as
\begin{equation}
\delta_P - \delta_S = 2 \arctan \frac{\beta}{\sqrt{\alpha^2+\beta^2}+\alpha},
\label{eq:phase}
\end{equation}
in terms of the measured Lee-Yang parameters $\alpha$ and $\beta$. %Specifically,  $\alpha$ and $\beta$ 
%represent corresponding measurements in experiment
\footnote{
The phase-shift formula presented by BESIII \cite{BESIII:2023wrw}, 
$\delta_P - \delta_S = \arctan({\beta}/{\alpha})$, is somewhat misleading 
as the range of their solution is limited to $(-\frac{\pi}{2}, \frac{\pi}{2})$, which does not 
fully cover the phase-shift space. 
However, in practical simulations, BESIII's solution manages to cover the full space through manual adjustment, acknowledging the formula's inherent limitations.
In contrast, Eq. (\ref{eq:phase}) proposed here naturally 
covers the correct solution space without imposing manual modification.
}
%The  phase shift formula presented by the BESIII \cite{BESIII:2023wrw},  
%$\delta_P - \delta_S = \arctan({\beta}/{\alpha})$,  is somewhat misleading as it
%yields two solutions for the  phase shift in the $\Lambda_c^+\to \Xi^0 K^+$ decay.
%By employing Eq. (\ref{eq:phase}) together with the measured values $\alpha=0.01\pm 0.16$ and $\beta=-0.64\pm 0.70$,
%we derive one unique solution of $\delta_P - \delta_S = -1.56 \pm 0.27$ rad. Consequently,  this suggests  that only one of the two %solutions 
% provided by the BESIII, namely, $\delta_P - \delta_S = -1.55 \pm 0.25 \pm 0.05$ rad,  is valid, while the other solution with $1.59\pm0.25\pm0.05$ rad should be abandoned.
In theoretical calculations, they are denoted by ${\textrm{Re}}(A^*B)$ and ${\textrm{Im}}(A^*B)$ terms, respectively.
It is easily seen that $\alpha^2+\beta^2+\gamma^2=1$.
The available experimental data are collected in the Appendix, i.e. Table \ref{tab:fitCF} below. Note that for $\Xi_c^{0}$ decays, several modes are measured relative to $\Xi_c^0\to \Xi^-\pi^+$; that is, ${\cal R}_X\equiv \B(\Xi_c^0\to X)/\B(\Xi_c^0\to \Xi^-\pi^+)$ for $X=\Xi^-K^+, \Lambda^0K_S^0, \Sigma^0K_S^0$ and $\Sigma^+K^-$. 
To compute the branching fractions, we need the inputs from 
the charmed baryon lifetimes which we shall use \cite{Cheng:2023jpz,Workman:2022ynf}
\begin{equation}
\label{eq:lifetimeWA}
\tau(\Xi_c^+)=(453\pm 5) ~{\rm fs}, \qquad \tau(\Lambda_c^+)=(202.9\pm 1.1) ~{\rm fs}, \qquad \tau(\Xi_c^0)=(150.5\pm 1.9) ~{\rm fs}.
\end{equation}

In practice, 
we shall make use of the package \texttt{iminuit} \cite{iminuit, James:1975dr}
to search for $\chi^2_{\text{min}}$ together with
its corresponding fitting parameters $c_i$, and generate the covariance matrix 
among  parameters which further helps 
predict physical observables. The fitting 19 TDA and IRA parameters given in Eqs. (\ref{eq:19parameters}) and (\ref{eq:IRA19parameters}), respectively, are exhibited in Table \ref{tab:coeff}. Note that our results for the IRA amplitudes $\tilde{f}^x$ are numerically different from that given in Eq. (9) of Ref. \cite{Geng:2023pkr}. This may be partially ascribed to the fact that we use 30 instead of 29 data points for fit and employ the Belle's result $(1.80\pm0.52)\%$ \cite{Belle:2018kzz} for $\B(\Xi_c^0\to\Xi^-\pi^+)$ rather than the PDG value of $(1.43\pm0.32)\%$ \cite{Workman:2022ynf}. Nevertheless, our predicted branching fractions, decay asymmetries and phase shifts are in gross agreement with Ref. \cite{Geng:2023pkr}.

We see from Eq. (\ref{eq:kin}) that both the branching fraction and the longitudinal decay asymmetry $\alpha$ remain insensitive to the sign of the phase shift. Consequently, the current global fit, which incorporates 30 experimental inputs including solely branching fractions and the decay asymmetry $\alpha$, lacks the ability to discern phase-shift signs. In contrast, the transverse asymmetry $\beta$, being proportional to the sine of the phase shift, aids in determining the sign of $\delta_P-\delta_S$. Although the measurement of $\beta$ has been carried out in the $\Lambda_c^+ \to \Xi^0 K^+$ channel \cite{BESIII:2023wrw}, a determination of the phase-shift sign proves challenging owing to its considerable uncertainty, which aligns with the presence of two phase-shift solutions at the current stage (see Eq. (\ref{eq:phaseshifts}) below). In our numerical analysis,
we indeed obtain two sets of solutions, distinct from a global sign difference. These solutions await a further discrimination by the forthcoming 
$\beta$ measurements in the near future.

To obtain the numerical values in Table \ref{tab:coeff}, we have set $\delta^{\tilde{T}}_S=0$
and $\delta^{\tilde{f^a}}_S=0$. We notice that no matter which phase $\delta^{{X_i}}_S$ or
$\delta^{{X_i}}_P$ is set to zero, the phase difference $\delta_P^{X_i}-\delta_S^{X_i}$ remains the same except for a sign ambiguity. Indeed, this is the so-called $Z_2$ ambiguity in Ref. \cite{Geng:2023pkr}, namely, $(\delta_S^{X_i}, \delta_P^{X_i})\to (-\delta_S^{X_i}, -\delta_P^{X_i})$. Since $\beta$ is proportional to $\sin(\delta_P-\delta_S)$, this sign ambiguity can be resolved by the measurement of $\beta$ as just noticed in passing.

Since the phase differences $\delta_P-\delta_S$ for $\tilde {E}_1$ and $\tilde f^c$ are $-2.14\pm0.57$ and $-2.62\pm 1.55$ rad, respectively, it is clear from Table \ref{tab:coeff} that the magnitudes of $S$-wave or $P$-wave component of $\tilde{E}_1$ and $\tilde f^c$ are consistent with each other within errors. Hence, the relation $\tilde{E}_1=\tilde{f}^c$ is numerically satisfied within errors. Likewise, the relation $\tilde{E}_h=\tilde{f}^a$ is also verified. 

In the TDA, there are many modes receive contributions only from one of the tidle TDA amplitudes:

\begin{enumerate}
    \item $\tilde{T}$: $\Xi_c^0\to \Xi^-\pi^+, \Sigma^-\pi^+,\Xi^- K^+, \Sigma^-K^+$;~ $\Xi_c^+\to\Xi^0K^+$.
    \item $\tilde{C}$: $\Lambda_c^+\to p\overline{K}^0$;~ $\Xi_c^0\to \Xi^0 K^0$;~ $\Xi_c^+\to \Sigma^+ K^0$.
    \item $\tilde{C'}$: $\Lambda_c^+\to \Sigma^+K^0, \Sigma^0 K^+$;~ $\Xi_c^+\to p \overline{K}^0$.
    \item $\tilde{E}_1$: $\Lambda_c^+\to \Xi^0K^+$;~ $\Xi_c^0\to \Sigma^+ K^-, \Sigma^+\pi^-, p K^-, p\pi^-, n\pi^0$;~ $\Xi_c^+\to p \pi^0, n\pi^+$. 
\end{enumerate}
For example, both CF decays $\Lambda_c^+\to \Xi^0K^+$ and $\Xi_c^0\to \Sigma^+ K^-$ which have been observed proceed solely 
via the $W$-exchange diagrams characterized by the topological amplitude $\tilde{E}_1$.

\begin{table*}[t]
\caption{
The fit results based on the tilde TDA (upper entry) and tilde IRA (lower entry). $S$- and $P$-wave amplitudes are in units of $10^{-2}G_F\,{\rm GeV}^2$ and $\delta_P-\delta_S$ in radian. Experimental results are taken from Table \ref{tab:expandave}.
}
\label{tab:fitCF}
\centering
\resizebox{\textwidth}{!} 
{
\begin{tabular}
{ l |c rrr rrr |  c c}
\hline
\hline
Channel
&$10^{2}\mathcal{B}$
&$\alpha$~~~~~~~&$\beta$~~~~~~~&$\gamma$~~~~~~
&$|A|$~~~~&$|B|$~~~~ &$\delta_P-\delta_S$~~~ 
&$\mathcal{B}_\text{exp}$
&$\alpha_\text{exp}$\\
\hline
\multirow{2}{*}{$\Lambda_c^+\to\Lambda^0\pi^+$}&$1.31\pm0.05$&$-0.76\pm0.01$&$-0.17\pm0.24$&$-0.63\pm0.06$&$2.76\pm0.25$&$16.96\pm0.39$&$-2.92\pm0.30$&\multirow{2}{*}{$1.29\pm{0.05}$}&\multirow{2}{*}{$-0.76\pm0.01$}\\&$1.31\pm0.05$&$-0.76\pm0.01$&$-0.28\pm0.33$&$-0.59\pm0.16$&$2.91\pm0.57$&$16.78\pm0.82$&$-2.79\pm0.39$\\
\multirow{2}{*}{$\Lambda_c^+\to\Sigma^0\pi^+$}&$1.26\pm0.05$&$-0.48\pm0.02$&$0.86\pm0.07$&$-0.17\pm0.35$&$4.07\pm0.86$&$15.48\pm2.30$&$2.08\pm0.04$&\multirow{2}{*}{$1.27\pm{0.06}$}&\multirow{2}{*}{$-0.47\pm0.03$}\\&$1.25\pm0.05$&$-0.48\pm0.02$&$0.79\pm0.23$&$-0.39\pm0.47$&$3.49\pm1.35$&$16.81\pm2.85$&$2.11\pm0.13$\\
\multirow{2}{*}{$\Lambda_c^+\to\Sigma^+ \pi^0$}&$1.27\pm0.05$&$-0.48\pm0.02$&$0.86\pm0.07$&$-0.17\pm0.35$&$4.07\pm0.86$&$15.48\pm2.30$&$2.08\pm0.04$&\multirow{2}{*}{$1.25\pm0.09$}&\multirow{2}{*}{$-0.49\pm0.03$}\\&$1.26\pm0.05$&$-0.48\pm0.02$&$0.79\pm0.23$&$-0.39\pm0.47$&$3.49\pm1.35$&$16.81\pm2.85$&$2.11\pm0.13$\\
\multirow{2}{*}{$\Lambda_c^+\to\Sigma^+ \eta$}&$0.33\pm0.04$&$-0.93\pm0.04$&$-0.34\pm0.15$&$-0.13\pm0.24$&$2.30\pm0.35$&$9.48\pm1.12$&$-2.80\pm0.15$&\multirow{2}{*}{$0.32\pm0.04$}&\multirow{2}{*}{$-0.99\pm0.06$}\\&$0.31\pm0.04$&$-0.95\pm0.05$&$-0.30\pm0.15$&$0.09\pm0.29$&$2.51\pm0.39$&$8.29\pm1.35$&$-2.83\pm0.16$\\
\multirow{2}{*}{$\Lambda_c^+\to\Sigma^+ \eta'$}&$0.39\pm0.12$&$-0.45\pm0.07$&$0.89\pm0.04$&$0.03\pm0.51$&$3.81\pm1.45$&$23.04\pm3.76$&$2.03\pm0.08$&\multirow{2}{*}{$0.44\pm0.15$}&\multirow{2}{*}{$-0.46\pm0.07$}\\&$0.41\pm0.13$&$-0.46\pm0.07$&$0.80\pm0.39$&$-0.38\pm0.82$&$3.03\pm2.33$&$28.15\pm6.48$&$2.09\pm0.22$\\
\multirow{2}{*}{$\Lambda_c^+\to\Xi^0 K^+$}&$0.41\pm0.03$&$-0.16\pm0.13$&$-0.24\pm0.28$&$0.96\pm0.07$&$3.89\pm0.19$&$2.43\pm2.12$&$-2.15\pm0.65$&\multirow{2}{*}{$0.55\pm0.07$}&\multirow{2}{*}{$0.01\pm0.16$}\\&$0.42\pm0.03$&$-0.19\pm0.12$&$-0.11\pm0.42$&$0.98\pm0.05$&$3.99\pm0.18$&$1.85\pm2.10$&$-2.62\pm1.68$\\
\multirow{2}{*}{$\Lambda_c^+\to\Lambda^0 K^+$}&$0.0639\pm0.0030$&$-0.56\pm0.05$&$0.83\pm0.04$&$0.04\pm0.35$&$1.09\pm0.18$&$3.32\pm0.59$&$2.17\pm0.06$&\multirow{2}{*}{$0.0635\pm0.0031$}&\multirow{2}{*}{$-0.585\pm0.052$}\\&$0.0639\pm0.0030$&$-0.57\pm0.05$&$0.82\pm0.08$&$-0.11\pm0.50$&$1.00\pm0.29$&$3.57\pm0.80$&$2.18\pm0.08$\\
\multirow{2}{*}{$\Lambda_c^+\to\Sigma^0 K^+$}&$0.0376\pm0.0032$&$-0.54\pm0.08$&$0.35\pm0.34$&$-0.76\pm0.17$&$0.40\pm0.15$&$3.86\pm0.26$&$2.56\pm0.44$&\multirow{2}{*}{$0.0382\pm0.0051$}&\multirow{2}{*}{$-0.55\pm0.20$}\\&$0.0388\pm0.0032$&$-0.54\pm0.09$&$0.11\pm0.56$&$-0.83\pm0.11$&$0.34\pm0.11$&$4.00\pm0.21$&$2.95\pm0.98$\\
\multirow{2}{*}{$\Lambda_c^+\to\Sigma^+K_S$}&$0.0377\pm0.0032$&$-0.54\pm0.08$&$0.35\pm0.34$&$-0.77\pm0.17$&$0.40\pm0.15$&$3.86\pm0.26$&$2.56\pm0.44$&\multirow{2}{*}{$0.047\pm0.014$}&\\&$0.0389\pm0.0032$&$-0.54\pm0.09$&$0.11\pm0.56$&$-0.83\pm0.11$&$0.34\pm0.11$&$4.00\pm0.21$&$2.95\pm0.98$\\
\multirow{2}{*}{$\Lambda_c^+\to n\pi^+$}&$0.063\pm0.009$&$-0.78\pm0.13$&$-0.62\pm0.16$&$0.00\pm0.26$&$1.01\pm0.14$&$2.43\pm0.39$&$-2.47\pm0.30$&\multirow{2}{*}{$0.066\pm0.013$}&\\&$0.059\pm0.008$&$-0.81\pm0.14$&$-0.57\pm0.23$&$0.12\pm0.37$&$1.03\pm0.17$&$2.19\pm0.53$&$-2.52\pm0.27$\\
\multirow{2}{*}{$\Lambda_c^+\to p\pi^0$}&$0.0176\pm0.0032$&$-0.11\pm0.69$&$-0.88\pm0.29$&$0.46\pm0.63$&$0.64\pm0.12$&$0.94\pm0.59$&$-1.69\pm0.76$&\multirow{2}{*}{$0.0156_{-0.0061}^{+0.0075}$}&\\&$0.0208\pm0.0045$&$-0.69\pm0.31$&$-0.61\pm0.54$&$-0.40\pm0.75$&$0.45\pm0.25$&$1.64\pm0.57$&$-2.42\pm0.58$\\
\multirow{2}{*}{$\Lambda_c^+\to p K_S$}&$1.55\pm0.06$&$0.01\pm0.24$&$0.37\pm0.34$&$-0.93\pm0.13$&$1.41\pm1.34$&$18.68\pm0.71$&$1.54\pm0.65$&\multirow{2}{*}{$1.59\pm0.07$}&\multirow{2}{*}{$0.18\pm0.45$}\\&$1.57\pm0.06$&$0.03\pm0.24$&$0.42\pm0.41$&$-0.91\pm0.19$&$1.64\pm1.67$&$18.69\pm1.00$&$1.50\pm0.54$\\
\multirow{2}{*}{$\Lambda_c^+\to p\eta$}&$0.151\pm0.008$&$0.07\pm0.30$&$0.77\pm0.26$&$-0.63\pm0.33$&$1.01\pm0.44$&$5.46\pm0.57$&$1.48\pm0.38$&\multirow{2}{*}{$0.149\pm0.008$}&\\&$0.149\pm0.008$&$0.36\pm0.29$&$0.75\pm0.28$&$-0.56\pm0.44$&$1.09\pm0.55$&$5.31\pm0.75$&$1.12\pm0.32$\\
\multirow{2}{*}{$\Lambda_c^+\to p\eta'$}&$0.052\pm0.008$&$-0.54\pm0.19$&$0.62\pm0.19$&$-0.56\pm0.35$&$0.77\pm0.30$&$4.72\pm0.73$&$2.29\pm0.13$&\multirow{2}{*}{$0.049\pm0.009$}&\\&$0.053\pm0.009$&$-0.01\pm0.37$&$0.96\pm0.08$&$-0.26\pm0.64$&$1.01\pm0.43$&$4.28\pm1.22$&$1.59\pm0.38$\\
\multirow{2}{*}{$\Xi_c^0\to\Xi^- \pi^+$}&$2.83\pm0.10$&$-0.72\pm0.03$&$0.29\pm0.27$&$-0.63\pm0.13$&$4.51\pm0.79$&$31.47\pm1.31$&$2.76\pm0.32$&\multirow{2}{*}{$1.80\pm0.52$}&\multirow{2}{*}{$-0.64\pm0.05$}\\&$2.87\pm0.10$&$-0.72\pm0.03$&$0.13\pm0.45$&$-0.68\pm0.10$&$4.21\pm0.64$&$32.19\pm1.07$&$2.96\pm0.60$\\
\multirow{2}{*}{$\Xi_c^+\to\Xi^0\pi^+$}&$0.9\pm0.2$&$-0.93\pm0.07$&$0.35\pm0.20$&$-0.09\pm0.22$&$2.27\pm0.31$&$8.21\pm1.17$&$2.79\pm0.23$&\multirow{2}{*}{$1.6\pm0.8$}&\\&$0.8\pm0.1$&$-0.93\pm0.09$&$0.35\pm0.28$&$-0.14\pm0.28$&$2.12\pm0.33$&$8.05\pm1.31$&$2.78\pm0.29$\\
\hline
Channel
&$10^{2}\mathcal{R}_X$&$\alpha$~~~~~~~&$\beta$~~~~~~~&$\gamma$~~~~~~
&$|A|$~~~~&$|B|$~~~~ &$\delta_P-\delta_S$~~~ 
&$10^{2}(\mathcal{R}_X)_\text{exp}$
&$\alpha_\text{exp}$\\
\hline
\multirow{2}{*}{$\Xi_c^0\to\Xi^-K^+$}&$4.10\pm0.05$&$-0.76\pm0.03$&$0.31\pm0.28$&$-0.58\pm0.14$&$1.04\pm0.18$&$7.25\pm0.30$&$2.76\pm0.32$&\multirow{2}{*}{$2.75\pm0.57$}&\\&$4.08\pm0.04$&$-0.76\pm0.03$&$0.14\pm0.47$&$-0.63\pm0.11$&$0.97\pm0.15$&$7.41\pm0.25$&$2.96\pm0.60$\\
\multirow{2}{*}{$\Xi_c^0\to\Lambda K_S^0$}&$24.1\pm1.0$&$-0.23\pm0.16$&$0.69\pm0.26$&$-0.69\pm0.25$&$2.06\pm0.84$&$13.93\pm1.17$&$1.89\pm0.25$&\multirow{2}{*}{$22.9\pm1.4$}&\\&$24.1\pm0.9$&$-0.18\pm0.12$&$0.66\pm0.36$&$-0.73\pm0.31$&$1.95\pm1.12$&$14.21\pm1.45$&$1.83\pm0.26$\\
\multirow{2}{*}{$\Xi_c^0\to\Sigma^0 K_S^0$}&$4.0\pm0.6$&$0.01\pm0.61$&$-0.78\pm0.37$&$0.49\pm0.51$&$1.92\pm0.40$&$3.47\pm1.83$&$-1.56\pm0.70$&\multirow{2}{*}{$3.8\pm0.7$}&\\&$3.8\pm0.6$&$-0.63\pm0.40$&$-0.65\pm0.53$&$-0.28\pm0.83$&$1.32\pm0.80$&$5.38\pm1.66$&$-2.29\pm0.62$\\
\multirow{2}{*}{$\Xi_c^0\to\Sigma^+K^-$}&$13.0\pm1.1$&$-0.21\pm0.17$&$-0.33\pm0.37$&$0.92\pm0.14$&$3.89\pm0.19$&$2.43\pm2.12$&$-2.15\pm0.65$&\multirow{2}{*}{$12.3\pm1.2$}&\\&$13.3\pm1.1$&$-0.26\pm0.16$&$-0.15\pm0.57$&$0.96\pm0.10$&$3.99\pm0.18$&$1.85\pm2.10$&$-2.62\pm1.68$\\
\hline
\hline
\end{tabular}
}
\end{table*}

The predicted branching fractions, Lee-Yang parameters, the magnitudes of $S$ and $P$ waves and their phase shifts based on the TDA and IRA are shown in Tables  \ref{tab:fitCF}-\ref{tab:fitother2}. In general, the predictions based on both the TDA and IRA agree with each other as it should be except for a few discrepancies, for example, in $\Lambda_c^+\to p\pi^0$, $\Xi_c^+\to \Sigma^+\eta^{(')}$ and $\Xi_c^+\to \Sigma^+ K_S$. 
We see that the fitted results for the branching fractions and decay asymmetries are in good agreement with experiment except for the following three modes: $\Xi_c^0\to \Xi^-\pi^+$, $\Lambda_c^+\to\Xi^0 K^+$ and the ratio ${\cal R}_{\Xi^-K^+}$. The $\chi^2$ value of our fit is 2.0 per degree of freedom.
Our predicted 
branching fraction of $(2.83\pm0.10)\%$ for
$\Xi_c^0\to \Xi^-\pi^+$ is noticeably higher than the measured value of $(1.80\pm0.52)\%$ by Belle \cite{Belle:2018kzz}. A similar result of $(2.72\pm0.09)\%$ has also been obtained in Ref. \cite{Geng:2023pkr}. 
Using the first sum-rule derived in Eq. (\ref{eq:sumrule}) and the measured data collected in Table \ref{tab:expandave}, we find $\B(\Xi_c^0\to \Xi^-\pi^+)=(2.85\pm0.30)\%$ in excellent agreement with the aforementioned prediction.

\begin{table*}[h!]\footnotesize
%\begin{table}%\footnotesize[h]
\caption{
Same as Table \ref{tab:fitCF} except for yet-observed CF and SCS  modes.
}
\label{tab:fitother}
% \resizebox{\textwidth}{!} 
% {
\centering
\begin{tabular}
{ l | r rrr rr r
% | l|r rrr rr r
}
\hline
\hline
Channel&
$10^{3}\mathcal{B}$~~~~ &$\alpha$~~~~~~~&$\beta$~~~~~~~&$\gamma$~~~~~~
&$|A|$~~~ & $|B|$~~~ & $\delta_P-\delta_S$~~~\\
\hline
\multirow{2}{*}{$\Lambda_c^{+} \rightarrow p K_L$}&$15.37\pm0.62$&$-0.03\pm0.22$&$0.37\pm0.33$&$-0.93\pm0.13$&$1.38\pm1.27$&$18.48\pm0.71$&$1.65\pm0.62$\\&$15.49\pm0.65$&$-0.02\pm0.22$&$0.41\pm0.41$&$-0.91\pm0.18$&$1.54\pm1.60$&$18.47\pm0.97$&$1.62\pm0.55$\\
\multirow{2}{*}{$\Xi_c^{+} \rightarrow \Sigma^{+} K_S$}&$2.08\pm2.12$&$0.94\pm0.22$&$-0.17\pm0.80$&$0.28\pm1.15$&$
1.39\pm0.92$&$3.19\pm1.80$&$-0.18\pm0.84$\\&$4.77\pm3.82$&$0.88\pm
0.14$&$-0.42\pm0.45$&$-0.24\pm0.55$&$1.62\pm0.90$&$6.35\pm2.81$&$
-0.44\pm0.47$\\
\multirow{2}{*}{$\Xi_c^{+} \rightarrow p K_{S / L}$}&$2.00\pm0.20$&$-0.38\pm0.07$&$0.25\pm0.25$&$-0.89\pm0.09$&$0.40\pm0.15$&$3.86\pm0.26$&$2.56\pm0.44$\\&$2.10\pm0.19$&$-0.38\pm0.07$&$0.07\pm0.39$&$-0.92\pm0.05$&$0.34\pm0.11$&$4.00\pm0.21$&$2.95\pm0.98$\\
\multirow{2}{*}{$\Xi_c^{+} \rightarrow \Sigma^{+} \pi^0$}&$2.16\pm0.20$&$-0.07\pm0.30$&$0.93\pm0.14$&$-0.35\pm0.37$&$0.96\pm0.29$&$3.96\pm0.51$&$1.64\pm0.32$\\&$2.32\pm0.27$&$0.12\pm0.20$&$0.94\pm0.16$&$-0.33\pm0.48$&$1.01\pm0.37$&$4.07\pm0.73$&$1.44\pm0.21$\\
\multirow{2}{*}{$\Xi_c^{+} \rightarrow \Sigma^{+} \eta$}&$0.75\pm0.26$&$-0.02\pm0.57$&$-0.64\pm0.43$&$-0.77\pm0.35$&$0.36\pm0.24$&$3.09\pm0.76$&$-1.60\pm0.89$\\&$1.09\pm0.47$&$-0.01\pm0.56$&$-0.23\pm0.64$&$-0.97\pm0.15$&$0.15\pm0.39$&$3.95\pm0.96$&$-1.60\pm2.40$\\
\multirow{2}{*}{$\Xi_c^{+} \rightarrow \Sigma^{+} \eta'$}&$1.19\pm0.21$&$-0.31\pm0.11$&$0.92\pm0.10$&$-0.24\pm0.47$&$0.99\pm0.35$&$5.27\pm0.94$&$1.90\pm0.10$\\&$1.31\pm0.29$&$-0.32\pm0.13$&$0.81\pm0.37$&$-0.49\pm0.61$&$0.85\pm0.52$&$6.07\pm1.41$&$1.95\pm0.21$\\
\multirow{2}{*}{$\Xi_c^{+} \rightarrow \Sigma^0 \pi^{+}$}&$3.12\pm0.13$&$-0.59\pm0.04$&$0.72\pm0.13$&$-0.36\pm0.28$&$1.13\pm0.24$&$4.80\pm0.56$&$2.26\pm0.08$\\&$2.89\pm0.21$&$-0.56\pm0.04$&$0.69\pm0.28$&$-0.46\pm0.40$&$1.01\pm0.36$&$4.78\pm0.75$&$2.26\pm0.22$\\
\multirow{2}{*}{$\Xi_c^{+} \rightarrow \Xi^0 K^{+}$}&$1.00\pm0.16$&$-0.73\pm0.12$&$-0.57\pm0.17$&$0.38\pm0.22$&$1.01\pm0.14$&$2.43\pm0.39$&$-2.47\pm0.21$\\&$1.51\pm0.62$&$-0.62\pm0.31$&$-0.29\pm1.10$&$0.73\pm0.23$&$1.38\pm0.32$&$1.98\pm0.86$&$-2.70\pm1.62$\\
\multirow{2}{*}{$\Xi_c^0 \rightarrow \Sigma^0 K_L$}&$1.24\pm0.19$&$-0.20\pm0.61$&$-0.63\pm0.41$&$0.75\pm0.43$&$2.02\pm0.33$&$2.35\pm1.97$&$-1.88\pm1.02$\\&$1.87\pm0.44$&$-0.74\pm1.01$&$-0.49\pm1.04$&$0.47\pm0.54$&$2.31\pm0.44$&$4.29\pm2.27$&$-2.56\pm1.61$\\
\multirow{2}{*}{$\Xi_c^0 \rightarrow \Xi^0 \pi^0$}&$7.45\pm0.64$&$-0.51\pm0.08$&$0.34\pm0.33$&$-0.79\pm0.15$&$1.74\pm0.64$&$16.78\pm1.11$&$2.56\pm0.44$\\&$7.72\pm0.65$&$-0.51\pm0.09$&$0.10\pm0.53$&$-0.85\pm0.10$&$1.49\pm0.47$&$17.37\pm0.93$&$2.95\pm0.98$\\
\multirow{2}{*}{$\Xi_c^0 \rightarrow \Xi^0 \eta$}&$2.87\pm0.66$&$0.08\pm0.20$&$0.86\pm0.18$&$0.50\pm0.30$&$3.12\pm0.45$&$6.61\pm2.16$&$1.48\pm0.24$\\&$2.28\pm0.53$&$0.24\pm0.24$&$0.86\pm0.24$&$0.45\pm0.44$&$2.73\pm0.55$&$6.20\pm2.51$&$1.30\pm0.28$\\
\multirow{2}{*}{$\Xi_c^0 \rightarrow \Xi^0 \eta'$}&$5.31\pm1.33$&$-0.59\pm0.08$&$0.79\pm0.07$&$0.18\pm0.41$&$4.87\pm1.38$&$23.13\pm3.82$&$2.22\pm0.08$\\
&$5.66\pm1.62$&$-0.59\pm0.09$&$0.79\pm0.20$&$-0.16\pm0.71$&$4.24\pm2.23$&$28.35\pm6.88$&$2.21\pm0.19$\\
\multirow{2}{*}{$\Xi_c^0 \rightarrow \Lambda^0 K_L$}&$7.17\pm0.24$&$-0.27\pm0.14$&$0.72\pm0.24$&$-0.64\pm0.27$&$2.36\pm0.86$&$14.13\pm1.26$&$1.92\pm0.21$\\&$7.29\pm0.26$&$-0.22\pm0.10$&$0.69\pm0.34$&$-0.69\pm0.33$&$2.18\pm1.16$&$14.49\pm1.55$&$1.88\pm0.23$\\
\multirow{2}{*}{$\Xi_c^0 \rightarrow \Sigma^{+} \pi^{-}$}&$0.22\pm0.02$&$-0.23\pm0.18$&$-0.35\pm0.39$&$0.91\pm0.15$&$0.90\pm0.04$&$0.56\pm0.49$&$-2.15\pm0.65$\\&$0.22\pm0.02$&$-0.27\pm0.17$&$-0.16\pm0.61$&$0.95\pm0.11$&$0.92\pm0.04$&$0.43\pm0.48$&$-2.62\pm1.68$\\
\multirow{2}{*}{$\Xi_c^0 \rightarrow \Sigma^0 \pi^0 $}&$0.34\pm0.04$&$-0.02\pm0.24$&$-0.38\pm0.37$&$-0.93\pm0.15$&$0.22\pm0.22$&$3.26\pm0.23$&$-1.62\pm0.64$\\&$0.36\pm0.05$&$-0.20\pm0.26$&$-0.31\pm0.44$&$-0.93\pm0.17$&$0.22\pm0.27$&$3.34\pm0.32$&$-2.13\pm0.75$\\
\multirow{2}{*}{$\Xi_c^0 \rightarrow \Sigma^0 \eta$}&$0.12\pm0.04$&$-0.02\pm0.57$&$-0.64\pm0.43$&$-0.77\pm0.36$&$0.25\pm0.17$&$2.18\pm0.54$&$-1.60\pm0.89$\\&$0.18\pm0.08$&$-0.01\pm0.57$&$-0.23\pm0.64$&$-0.97\pm0.15$&$0.11\pm0.28$&$2.79\pm0.68$&$-1.60\pm2.40$\\
\multirow{2}{*}{$\Xi_c^0 \rightarrow \Sigma^0 \eta'$}&$0.20\pm0.04$&$-0.31\pm0.11$&$0.92\pm0.10$&$-0.24\pm0.47$&$0.70\pm0.24$&$3.73\pm0.66$&$1.90\pm0.10$\\&$0.22\pm0.05$&$-0.32\pm0.13$&$0.81\pm0.37$&$-0.49\pm0.61$&$0.60\pm0.37$&$4.29\pm1.00$&$1.95\pm0.21$\\
\multirow{2}{*}{$\Xi_c^0 \rightarrow \Sigma^{-} \pi^{+}$}&$1.92\pm0.08$&$-0.65\pm0.03$&$0.26\pm0.25$&$-0.71\pm0.10$&$1.04\pm0.18$&$7.25\pm0.30$&$2.76\pm0.32$\\&$1.26\pm0.18$&$-0.77\pm0.06$&$0.19\pm0.51$&$-0.61\pm0.14$&$0.97\pm0.15$&$5.70\pm0.58$&$2.90\pm0.64$\\
\multirow{2}{*}{$\Xi_c^0 \rightarrow \Xi^0 K_{S / L}$}&$0.43\pm0.02$&$-0.48\pm0.03$&$0.87\pm0.04$&$-0.05\pm0.36$&$0.94\pm0.20$&$3.56\pm0.53$&$2.08\pm0.04$\\&$0.41\pm0.03$&$-0.50\pm0.04$&$0.82\pm0.19$&$-0.28\pm0.51$&$0.80\pm0.31$&$3.87\pm0.66$&$2.11\pm0.13$\\
\hline
\hline
\end{tabular}
% }
%\end{table}
\end{table*}

\begin{table*}\footnotesize
%\begin{table}%\footnotesize[h]
\caption{
Same as Table \ref{tab:fitCF} except for yet-observed SCS and DCS  modes.
}
\label{tab:fitother2}
% \resizebox{\textwidth}{!} 
% {
\centering
\begin{tabular}
{ l | r rrr c c r
% | l|r rrr c c r
}
\hline
\hline
Channel& $10^{4}\mathcal{B}$~~~~ &$\alpha$~~~~~~~&$\beta$~~~~~~~&$\gamma$~~~~~~
&$|A|$~~~ & $|B|$~~~ & $\delta_P-\delta_S$~~~\\
\hline
\multirow{2}{*}{$\Lambda_c^{+} \rightarrow n K^{+}$}&$~0.12\pm0.03$&$-0.88\pm0.45$&$0.33\pm0.22$&$-0.34\pm0.24$&$0.12\pm0.02$&$0.44\pm0.06$&$2.79\pm0.23$\\&$0.12\pm0.02$&$-0.86\pm0.08$&$0.33\pm0.28$&$-0.38\pm0.24$&$0.11\pm0.02$&$0.43\pm0.07$&$2.78\pm0.29$\\
\multirow{2}{*}{$\Xi_c^{+} \rightarrow \Lambda^0 \pi^{+}$}&$3.55\pm1.15$&$0.21\pm0.28$&$-0.62\pm0.29$&$-0.76\pm0.22$&$0.24\pm0.12$&$1.70\pm0.28$&$-1.25\pm0.48$\\&$3.94\pm1.31$&$0.11\pm0.28$&$-0.82\pm0.28$&$-0.56\pm0.42$&$0.34\pm0.16$&$1.69\pm0.39$&$-1.43\pm0.33$\\
\multirow{2}{*}{$\Xi_c^{+} \rightarrow n \pi^{+}$}&$0.34\pm0.04$&$-0.28\pm0.22$&$-0.43\pm0.47$&$0.86\pm0.24$&$0.21\pm0.01$&$0.13\pm0.11$&$-2.15\pm0.65$\\&$0.35\pm0.03$&$-0.34\pm0.21$&$-0.20\pm0.76$&$0.92\pm0.18$&$0.21\pm0.01$&$0.10\pm0.11$&$-2.62\pm1.68$\\
\multirow{2}{*}{$\Xi_c^{+} \rightarrow \Sigma^0 K^{+}$}&$1.22\pm0.05$&$-0.68\pm0.03$&$0.28\pm0.26$&$-0.67\pm0.12$&$0.17\pm0.03$&$1.18\pm0.05$&$2.76\pm0.32$\\&$1.24\pm0.04$&$-0.68\pm0.03$&$0.13\pm0.42$&$-0.72\pm0.09$&$0.16\pm0.02$&$1.21\pm0.04$&$2.96\pm0.60$\\
\multirow{2}{*}{$\Xi_c^{+} \rightarrow p \pi^0$}&$0.17\pm0.02$&$-0.28\pm0.22$&$-0.43\pm0.47$&$0.86\pm0.24$&$0.15\pm0.01$&$0.09\pm0.08$&$-2.15\pm0.65$\\&$0.17\pm0.02$&$-0.34\pm0.21$&$-0.20\pm0.76$&$0.92\pm0.18$&$0.15\pm0.01$&$0.07\pm0.08$&$-2.62\pm1.68$\\
\multirow{2}{*}{$\Xi_c^{+} \rightarrow p \eta$}&$1.83\pm0.28$&$-0.40\pm0.07$&$0.58\pm0.21$&$-0.71\pm0.20$&$0.20\pm0.06$&$1.14\pm0.14$&$2.17\pm0.13$\\&$2.43\pm0.36$&$-0.34\pm 0.06$&$0.48\pm0.35$&$-0.81\pm0.22$&$0.18\pm0.10$&$1.35\pm0.17$&$2.19\pm0.32$\\
\multirow{2}{*}{$\Xi_c^{+} \rightarrow p \eta'$}&$0.99\pm0.17$&$-0.45\pm0.17$&$0.69\pm0.15$&$-0.56\pm0.31$&$0.20\pm0.07$&$1.06\pm0.17$&$2.15\pm0.12$\\&$1.24\pm0.43$&$-0.33\pm
0.21$&$0.54\pm0.38$&$-0.77\pm0.33$&$0.16\pm0.10$&$1.26\pm0.32$&$2.12\pm0.14$\\
\multirow{2}{*}{$\Xi_c^{+} \rightarrow \Lambda^0 K^{+}$}&$0.35\pm0.05$&$-0.41\pm0.13$&$0.72\pm0.24$&$0.56\pm0.31$&$0.20\pm0.02$&$0.30\pm0.12$&$2.09\pm0.20$\\&$0.35\pm0.05$&$-0.38\pm0.13$&$0.86\pm0.21$&$0.34\pm0.53$&$0.18\pm0.03$&$0.36\pm0.16$&$1.99\pm0.17$\\
\multirow{2}{*}{$\Xi_c^0 \rightarrow p K^{-}$}&$1.99\pm0.22$&$-0.27\pm0.21$&$-0.42\pm0.46$&$0.87\pm0.23$&$0.90\pm0.04$&$0.56\pm0.49$&$-2.15\pm0.65$\\&$2.03\pm0.18$&$-0.33\pm0.20$&$-0.19\pm0.74$&$0.92\pm0.17$&$0.92\pm0.04$&$0.43\pm0.48$&$-2.62\pm1.68$\\
\multirow{2}{*}{$\Xi_c^0 \rightarrow n K_{S / L}$}&$7.41\pm0.79$&$-0.43\pm0.05$&$0.78\pm0.14$&$-0.45\pm0.28$&$0.94\pm0.20$&$3.56\pm0.53$&$2.08\pm0.04$\\&$7.85\pm1.03$&$-0.40\pm0.05$&$0.67\pm0.28$&$-0.62\pm0.34$&$0.80\pm0.31$&$3.87\pm0.66$&$2.11\pm0.13$\\
\multirow{2}{*}{$\Xi_c^0 \rightarrow \Lambda^0 \pi^0 $}&$1.12\pm0.32$&$-0.61\pm0.20$&$-0.58\pm0.31$&$-0.55\pm0.27$&$0.31\pm0.12$&$1.54\pm0.22$&$-2.38\pm0.38$\\&$1.53\pm0.51$&$-0.26\pm0.45$&$-0.87\pm0.25$&$-0.42\pm0.38$&$0.42\pm0.16$&$1.73\pm0.35$&$-1.86\pm0.54$\\
\multirow{2}{*}{$\Xi_c^0 \rightarrow n \pi^0 $}&$0.06\pm0.01$&$-0.28\pm0.22$&$-0.43\pm0.47$&$0.86\pm0.24$&$0.15\pm0.01$&$0.09\pm0.08$&$-2.15\pm0.65$\\&$0.06\pm0.01$&$-0.34\pm0.21$&$-0.20\pm0.76$&$0.92\pm0.18$&$0.15\pm0.01$&$0.07\pm0.08$&$-2.62\pm1.68$\\
\multirow{2}{*}{$\Xi_c^0 \rightarrow \Lambda^0 \eta$}&$4.56\pm0.91$&$0.22\pm0.20$&$0.14\pm0.38$&$-0.97\pm0.08$&$0.19\pm0.19$&$4.00\pm0.44$&$0.57\pm1.27$\\&$4.81\pm1.17$&$0.03\pm0.22$&$0.06\pm0.51$&$-1.00\pm0.03$&$0.05\pm0.36$&$4.15\pm0.51$&$1.15\pm3.13$\\
\multirow{2}{*}{$\Xi_c^0 \rightarrow \Lambda^0 \eta'$}&$6.85\pm0.98$&$-0.64\pm0.11$&$0.71\pm0.10$&$-0.30\pm0.38$&$1.21\pm0.35$&$5.94\pm0.92$&$2.30\pm0.09$\\&$8.39\pm2.27$&$-0.51\pm0.15$&$0.63\pm0.35$&$-0.59\pm0.50$&$1.03\pm0.57$&$7.27\pm1.84$&$2.25\pm0.17$\\
\multirow{2}{*}{$\Xi_c^0 \rightarrow \Sigma^{-} K^{+}$}&$0.83\pm0.03$&$-0.68\pm0.03$&$0.28\pm0.26$&$-0.68\pm0.11$&$0.24\pm0.04$&$1.67\pm0.07$&$2.76\pm0.32$\\&$0.84\pm0.03$&$-0.68\pm0.03$&$0.13\pm0.42$&$-0.72\pm0.09$&$0.22\pm0.03$&$1.71\pm0.06$&$2.96\pm0.60$\\
\multirow{2}{*}{$\Xi_c^0 \rightarrow p \pi^{-}$}&$0.11\pm0.01$&$-0.28\pm0.22$&$-0.43\pm0.47$&$0.86\pm0.24$&$0.21\pm0.01$&$0.13\pm0.11$&$-2.15\pm0.65$\\&$0.12\pm0.01$&$-0.35\pm0.21$&$-0.20\pm0.76$&$0.92\pm0.18$&$0.21\pm0.01$&$0.10\pm0.11$&$-2.62\pm1.68$\\
\multirow{2}{*}{$\Xi_c^0 \rightarrow n \eta$}&$0.61\pm0.10$&$-0.40\pm0.07$&$0.58\pm0.21$&$-0.71\pm0.20$&$0.20\pm0.06$&$1.14\pm0.14$&$2.17\pm0.13$\\&$0.74\pm0.13$&$-0.37\pm0.08$&$0.41\pm0.37$&$-0.83\pm0.21$&$0.17\pm0.09$&$1.29\pm0.18$&$2.30\pm0.37$\\
\multirow{2}{*}{$\Xi_c^0 \rightarrow n \eta'$}&$0.33\pm0.06$&$-0.45\pm0.17$&$0.69\pm0.15$&$-0.56\pm0.31$&$0.20\pm0.07$&$1.06\pm0.17$&$2.15\pm0.12$\\&$0.41\pm0.14$&$-0.34\pm0.17$&$0.55\pm0.38$&$-0.76\pm0.34$&$0.17\pm0.10$&$1.25\pm0.31$&$2.12\pm0.14$\\
\hline
\hline
\end{tabular}
% }
%\end{table}
\end{table*}
% \end{widetext}

As for the ratio ${\cal R}_{\Xi^-K^+}$, we see from Table \ref{tab:CFSCSamp} that in the
SU(3) limit, one will have ${\cal R}_{\Xi^-K^+}=\sin^2\theta_C$ which is equal to 0.045 after taking into account the phase-space difference between  $\Xi_c^0\to \Xi^-K^+$ and $\Xi_c^0\to \Xi^-\pi^+$. The current measurement is $0.0275\pm0.0057$ which is away from the SU(3) expectation by $2\sigma$. Since both modes proceed through the topological diagrams $T$ and $E_{1S}$ with the combination $2T-2E_{1S}=2\tilde T$, it is conceivable that SU(3) breaking in the external $W$-emission $T$ and especially in $W$-exchange $E_{1S}$ could account for the discrepancy between theory and experiment. 

If the pseudoscalar meson in the final-state is an SU(3) flavor-singlet $\eta_1$, it will receive an additional contribution from the hairpin diagram $E_h$ in Fig. \ref{Fig:TopDiag}. Numerically, we find that the combination $\tilde E_h=E_h+2E_{1S}$
which contributes only to $\eta_1$ is sizable (see Table \ref{tab:coeff}). Recall that in the charmed meson sector there is a strong indication of the hairpin effect in the decay $D_s^+\to \rho^+\eta'$ (see e.g. Ref. \cite{Cheng:2024hdo}).

For the decay $\Lambda_c^+\to \Xi^0 K^+$, BESIII uncovered two sets of solutions for the magnitudes of $S$- and $P$-wave amplitudes in units of $10^{-2} G_F\,{\rm GeV}^2$:
\begin{equation}
\label{eq:BESIIISol}
{\rm I}.~
\begin{cases}
 |A|=1.6^{+1.9}_{-1.6}\pm0.4\,,  \\  |B|=18.3\pm2.8\pm0.7\,,
 \end{cases} 
 \qquad
{\rm II}.~ \begin{cases}
|A|=4.3^{+0.7}_{-0.2}\pm0.4\,, \\ |B|=6.7^{+8.3}_{-6.7}\pm1.6\,,
\end{cases}
\end{equation}
and two solutions for the phase shift, 
\begin{equation}
\label{eq:phaseshifts}
\delta_P-\delta_S=-1.55\pm 0.25\pm0.05~~{\rm or}~~1.59\pm0.25\pm0.05~{\rm rad}.
\end{equation}
Our fits with $|A|=3.89\pm0.19$,  $|B|=2.43\pm2.12$, $\alpha_{\Xi^0K^+}=-0.16\pm0.13$, $\beta_{\Xi^0K^+}=-0.24\pm0.28$ and 
$\delta_P-\delta_S=-2.15\pm 0.65$ rad are consistent with solution II for $|A|$ and $|B|$ and the first phase-shift solution as well as the Lee-Yang parameters $\alpha_{\Xi^0K^+}$ and $\beta_{\Xi^0K^+}$. 
However, our result of $\B(\Lambda_c^+\to \Xi^0 K^+)=(0.41\pm0.03)\%$ is slightly smaller than the measured value of $(0.55\pm0.07)\%$ \cite{Workman:2022ynf}.
Our results are very close to those obtained in Ref. \cite{Geng:2023pkr}:
$\B(\Lambda_c^+\to \Xi^0 K^+)=(0.40\pm0.03)\%$, $\alpha_{\Xi^0K^+}=-0.15\pm0.14$, $\beta_{\Xi^0K^+}=-0.29\pm0.22$ and $\delta_P-\delta_S=-2.06\pm 0.50$ rad.

It should be stressed that although the BESIII's measurement of $\alpha_{\Xi^0 K^+}$ is in good agreement with zero, it does not mean that the theoretical predictions in the 1990s with vanishing or very small $S$-wave amplitude are confirmed. 
We have checked that if we set $\delta_S^{X_i}=\delta_P^{X_i}=0$ from the outset and keep the measured $\alpha_{\Xi^0K^+}$ as an input, the fit $\B(\Lambda_c^+\to \Xi^0 K^+)$ of order $1\times 10^{-3}$ will be too small compared to experiment because of the smallness of the $S$-wave contribution. On the contrary, if the input of $(\alpha_{\Xi^0K^+})_{\rm exp}$ is removed, the fit $\alpha_{\Xi^0K^+}$ will be of order 0.95\,. Hence, we conclude that it is inevitable to incorporate the phase shifts to accommodate the data. It is the smallness of $|\cos(\delta_P-\delta_S)|\sim 0.02$ that accounts for the nearly vanishing  $\alpha_{\Xi^0K^+}$.

Besides the decay $\Lambda_c^+\to \Xi^0 K^+$, we have noticed in passing that  the following modes
$\Xi_c^0\to \Sigma^+ K^-, \Sigma^+\pi^-, p K^-, p\pi^-, n\pi^0$ and $\Xi_c^+\to p \pi^0, n\pi^+$ also receive contributions only from the topological $W$-exchange amplitude $\tilde{E}_1$. In the absence of strong phases in $S$- and $P$-wave amplitudes, they are expected to have large decay asymmetries. For example, $\alpha_{\Xi_c^0\to \Sigma^+K^-}$ was found to be $0.79^{+0.32}_{-0.33}$, $0.81\pm0.16$ and $0.98\pm0.20$, respectively, in Refs. \cite{Zhong:2022exp,Geng:2019xbo,Xing:2023dni}. Once the phase shifts are incorporated in the fit, the above-mentioned modes should have $\delta_P-\delta_S$ similar to that in $\Lambda_c^+\to \Xi^0 K^+$  and their decay asymmetries will become smaller (see Tables \ref{tab:fitCF}-\ref{tab:fitother2}).
In particular, for the CF channel $\Xi_c^0\to \Sigma^+ K^-$ whose branching fraction has been measured before, 
we predict that $\alpha_{\Xi_c^0\to \Sigma^+ K^-}=-0.21\pm0.17$ which can be used to test our theoretical framework.

It is worth emphasizing again that when the strong phases $\delta_S^{X_i}$ and $\delta_P^{X_i}$ are turned off for the global fit to the data, the decay asymmetry $\alpha$ is found to be close to unity in magnitude for some of the modes, e.g., $\Lambda_c^+\to \Sigma^+\eta,~\Xi^0 K^+,~\Sigma^0 K^+$; $\Xi_c^+\to \Xi^0\pi^+$
and $\Xi_c^0\to \Lambda^0 K_S^0, ~\Xi^0\eta$ \cite{Zhong:2022exp,Xing:2023dni}. As noticed in passing, the nearly vanishing $\alpha_{\Xi^0 K^+}$ measured by BESIII implies the necessity of incorporating phase shifts for global fits, which will also help explain the measured value of $\alpha_{\Sigma^0 K^+}$. The measurement of
$\alpha_{\Xi_c^0\to \Lambda K_S}$ and $\alpha_{\Xi_c^0\to \Xi^0\eta}$ in the future will also help understand the phase shifts which are predicted to be $1.89\pm0.25$ and $1.48\pm0.24$ (both in rad), respectively. In contrast, it is also important to measure the decay asymmetry of
$\Xi_c^+\to \Xi^0 \pi^+$ to see if its largeness of order 
$-0.93$ is not affected by the phase shift which is expected to be $\delta_P-\delta_S=2.79\pm0.23$ rad (see Table \ref{tab:fitCF}) and hence
$|\cos(\delta_P-\delta_S)|\sim 0.94$\,.

For the Cabibbo-favored (CF)  modes involving a neutral $K_S^0$ or $K_L^0$, it was customary to use the relation $\Gamma(\overline  K^0)=2\Gamma(K_S^0)$. However, this relation can be invalidated by the interference between CF and doubly Cabibbo-suppressed (DCS) amplitudes. Using the phase convention  $K_S^0={1\over\sqrt{2}}(K^0-\overline K^0)$ and $K_L^0={1\over\sqrt{2}}(K^0+\overline K^0)$ in the absence of $C\!P$ violation, we have
\begin{equation}
\begin{split}
A(\B_c\to \B K_S^0 ) &= -{1\over\sqrt{2}}[A(\B_c\to \B \overline K^0)-A(\B_c\to \B K^0 )],
\\
A(\B_c\to \B K_L^0 ) &= {1\over\sqrt{2}}[A(\B_c\to \B \overline K^0)+A(\B_c\to \B K^0 )].
\end{split}
\end{equation}
Since $\B_c\to \B K^0$ is doubly Cabibbo-suppressed, it is expected that $\Gamma(\B_c\to \B K_L^0 )\approx \Gamma(\B_c\to \B K_S^0)$. For singly Cabibbo-suppressed (SCS) channels, $\Xi_c^+\to pK_{S/L}$ and $\Xi_c^0\to nK_{S/L}$ receive contributions only from $\overline K^0$, while $\Xi_c^0\to \Xi^0 K_{S/L}$ proceeds only through $K^0$.

\section{Conclusion}
\label{sec:dis}
Inspired by the recent BESIII measurement of the decay asymmetry and the phase shift between $S$- and $P$-wave amplitudes in the decay $\Lambda_c^+\to \Xi^0K^+$, we have performed a global fit to the experimental data of charmed baryon decays based on the TDA which has the advantage that it is more intuitive, graphic and easier to implement model calculations. Our main results are as follows.

\begin{itemize}
\item
In order to draw the topological diagrams and construct the relevant amplitudes in the TDA, we employ the anti-symmetric matrix $(\B_c)^{ij}$ for the charmed baryon and $(B_8)_{ijk}$ for the octet baryon, where the indices $i,j,k$ stand for light quark flavors. 

\item       
The wave functions of octet baryons can be represented in several different manners, but physics is independent of the convention on chooses. We use the bases $\psi(8)_{A_{12}}$ and $\psi(8)_{S_{12}}$ to assign  different topological diagrams and amplitudes. After applying for the KPW theorem, the number of independent topological diagrams and amplitudes is reduced to 7. At this stage, there still exist 2 redundant degrees of freedom through redefinition. We conclude that the minimum set of the topological amplitudes in the TDA is 5.    

\item 
To demonstrate the equivalence between the TDA and IRA, we have shown that the number of the minimum set of tensor invariants in the IRA and the topological amplitudes in the TDA is the same and presented their relations. 

\item 
As there are 5 independent tilde TDA amplitudes,  we have totally 19 unknown parameters to describe the magnitudes and the phases of the respective $S$- and $P$-wave amplitudes. The measured branching fractions and decay asymmetries are well accommodated in the TDA except for three modes: $\Xi_c^0\to \Xi^-\pi^+$, $\Lambda_c^+\to\Xi^0 K^+$ and the ratio ${\cal R}_{\Xi^-K^+}$. The $\chi^2$ value is 2.0 per degree of freedom. 
    The predicted $\B(\Xi_c^0\to \Xi^-\pi^+)=(2.83\pm0.10)\%$ is larger than its current value, but it is in agood agreement with the sum rule derived in both the TDA and IRA.  This needs to be tested in the near future.

\item 
The phase difference $\delta_P-\delta_S$ between $S$- and $P$-wave amplitudes is subject to a sign ambiguity which can be 
  resolved by the measurement of the transverse asymmetry $\beta$. Hence, more accurate measurements of $\beta$ in $\Lambda_c^+\to \Xi^0K^+$ and other modes are called for. 

\item 
When the strong phases $\delta_S^{X_i}$ and $\delta_P^{X_i}$ are turned off for the global fit to the data, the decay asymmetry $\alpha$ is found to be close to unity in magnitude for some of the modes, e.g., $\Lambda_c^+\to \Sigma^+\eta,~\Xi^0 K^+,~\Sigma^0 K^+$, $\Xi_c^+\to \Xi^0\pi^+$
and $\Xi_c^0\to \Lambda^0 K_S^0, ~\Xi^0\eta$. The nearly vanishing $\alpha_{\Xi^0 K^+}$ measured by BESIII implies the necessity of incorporating phase shifts in order to accommodate the data. The measurement of
$\alpha_{\Xi_c^0\to \Lambda K_S}$ and $\alpha_{\Xi_c^0\to \Xi^0\eta}$ in the future will also help understand the phase shifts which are predicted to be $1.89\pm0.25$ and $1.48\pm0.24$ (both in rad), respectively. In contrast, it is also important to measure the decay asymmetry of
$\Xi_c^+\to \Xi^0 \pi^+$ to see if its largeness of order 
$-0.93$ is not affected even after the phase-shift effect is incorporated.

\item 
The fit results for the decay asymmetry and the phase shift $\delta_P-\delta_S$ for $\Lambda_c^+\to \Xi^0K^+$ are consistent with the BESIII measurements, though our fit of its branching fraction is slightly smaller than the measured one. Nevertheless, our TDA results are very close to those obtained in  the IRA. 

\item 
Besides the decay $\Lambda_c^+\to \Xi^0 K^+$, there exist several modes
$\Xi_c^0\to \Sigma^+ K^-, \Sigma^+\pi^-, p K^-, p\pi^-, n\pi^0$ and $\Xi_c^+\to p \pi^0, n\pi^+$ which proceed only from $W$-exchange characterized by the topological amplitude $\tilde{E}_1$. By the same token, these modes should have phase shifts similar to that in $\Lambda_c^+\to \Xi^0 K^+$  and their decay asymmetries are suppressed.  
In particular, we predict  $\alpha_{\Xi_c^0\to \Sigma^+ K^-}=-0.21\pm0.17$ which can be used to test our theoretical framework.

\item 
For yet-to-be-measured modes, we have presented the fitting magnitudes of $S$- and $P$-wave amplitudes and their phase shifts in both the TDA and IRA which can be tested in the near future. 

\end{itemize}

\begin{acknowledgments}
		We would like to thank Pei-Rong Li and Chia-Wei Liu for valuable discussions.
		This research was supported in part by the Ministry of Science and Technology of R.O.C. under Grant No. MOST-112-2112-M-001-026 and 
		the National Natural Science Foundation of China
		under Grant No. U1932104.
\end{acknowledgments}

%\newpage

\appendix
\section{Experimental Data}
Experimental data are collected in Table \ref{tab:expandave}.
In addition to the latest PDG values \cite{Workman:2022ynf} adopted as 
partial inputs in $\chi^2$,  
more $\Lambda_c^+$
related data have been supplied by BESIII in 2023, including branching fractions of 
$\Lambda_c^+\to p \pi^0$ \cite{BESIII:2023uvs} and 
$\Lambda_c^+ \to p \eta$ \cite{BESIII:2023wrw}.
On the other hand, Belle has contributed the recent measurements on $\Xi_c^{0}$, 
such as $\Xi_c^0\to \Xi^- \pi^+$ \cite{Belle:2018kzz} and $\Xi_c^0\to \Lambda^0 K_S, \Sigma^0 K_S, \Sigma^+ K^-$ \cite{Belle:2021avh}.

%----------Exprimental data----------
\begin{table}%[h]%\footnotesize
\caption{Experimental data of branching fractions and decay asymmetries taken from PDG, BESIII and Belle. The data marked with an asterisk have been included in the PDG average. } 
\vspace{-0.4cm}
\label{tab:expandave}
\begin{center}
\renewcommand\arraystretch{1}
\resizebox{\textwidth}{!} 
{
\begin{tabular}
{ l c c c c }
\hline
Observable & PDG \cite{Workman:2022ynf} & BESIII & Belle & Average\\
\hline
$10^{2}\mathcal{B}(\Lambda_c^+\to\Lambda^0\pi^+)$&
$1.29\pm{0.05}$&&&
$1.29\pm{0.05}$\\

$10^{2}\mathcal{B}(\Lambda_c^+\to \Sigma^0 \pi^+)$&
$1.27\pm{0.06}$&&&
$1.27\pm{0.06}$\\

$10^{2}\mathcal{B}(\Lambda_c^+\to \Sigma^+ \pi^0)$&
$1.25\pm0.09$&&&
$1.25\pm0.09$\\

$10^{2}\mathcal{B}(\Lambda_c^+\to \Sigma^+ \eta)$&
$0.44\pm0.20$&&$0.314\pm0.044$ \cite{Belle:2022bsi}&
$0.32\pm0.04$~\cite{Workman:2022ynf, Belle:2022bsi}\\

$10^{2}\mathcal{B}(\Lambda_c^+\to \Sigma^+ \eta')$&
$1.5\pm0.6$&&$0.416\pm0.086$~\cite{Belle:2022bsi}&
$0.44\pm0.15$~\cite{Workman:2022ynf, Belle:2022bsi}\\

$10^{2}\mathcal{B}(\Lambda_c^+\to \Xi^0 K^+)$&
$0.55\pm0.07$&&&
$0.55\pm0.07$\\

$10^{4}\mathcal{B}(\Lambda_c^+\to \Lambda^0 K^+)$&
$6.0\pm0.5$&$6.21\pm0.61^*$~\cite{BESIII:2022tnm}& $6.57\pm0.40$ ~\cite{Belle:2022uod}&
$6.35\pm0.31$~\cite{Workman:2022ynf,Belle:2022uod}\\

$10^{4}\mathcal{B}(\Lambda_c^+\to \Sigma^0 K^+)$&
$4.9\pm0.6$&$4.7\pm0.95^*$ ~\cite{BESIII:2022wxj}& $3.58\pm0.28$ ~\cite{Belle:2022uod}&
$3.82\pm0.51$ ~\cite{Workman:2022ynf,Belle:2022uod}\\

$10^{4}\mathcal{B}(\Lambda_c^+\to \Sigma^+ K_S)$&
$4.7\pm1.4$&$4.8\pm1.5^*$ ~\cite{BESIII:2022wxj} &&
$4.7\pm1.4$\\

$10^{4}\mathcal{B}(\Lambda_c^+\to n \pi^+)$&
$6.6\pm1.3$&&&
$6.6\pm1.3$\\

$10^{4}\mathcal{B}(\Lambda_c^+\to p\pi^0)$&
$<0.8$&$1.56_{-0.58}^{+0.72}\pm0.20$ ~\cite{BESIII:2023uvs}&&
$1.56_{-0.61}^{+0.75} $~\cite{BESIII:2023uvs}\\

$10^{2}\mathcal{B}(\Lambda_c^+\to p K_S)$&
$1.59\pm0.07$&&&
$1.59\pm0.07$\\

$10^{3}\mathcal{B}(\Lambda_c^+\to p \eta)$&
$1.41\pm0.11$ & $1.63\pm0.33$ ~\cite{BESIII:2023uvs}, $1.57\pm0.12$ ~\cite{BESIII:2023ooh}&&
$1.49\pm0.08$ ~\cite{Workman:2022ynf,BESIII:2023uvs,BESIII:2023ooh}\\

$10^{4}\mathcal{B}(\Lambda_c^+\to p \eta')$&
$4.9\pm0.9$&&&
$4.9\pm0.9$\\

$10^{2}\mathcal{B}(\Xi_c^0\to \Xi^- \pi^+)$&
$1.43\pm0.32$&&$1.80\pm0.52^*$ \cite{Belle:2018kzz}&
$1.80\pm0.52$ \cite{Belle:2018kzz}\\

$10^{2}\frac{\mathcal{B}(\Xi_c^0\to \Xi^- K^+)}{\mathcal{B}(\Xi_c^0\to \Xi^- \pi^+)}$&
$2.75\pm0.57$&&&$2.75\pm0.57$\\
%\cite{Belle:2013ntc}\\
$10^{2}\frac{\mathcal{B}(\Xi_c^0\to \Lambda K_S^0)}{\mathcal{B}(\Xi_c^0\to \Xi^- \pi^+)}$&
$22.5\pm1.3$&&$22.9\pm1.4^*$ \cite{Belle:2021avh}&$22.9\pm1.4$ \cite{Belle:2021avh}\\
$10^{2}\frac{\mathcal{B}(\Xi_c^0\to \Sigma^0 K_S^0)}{\mathcal{B}(\Xi_c^0\to \Xi^- \pi^+)}$&
$3.8\pm0.7$&&&$3.8\pm0.7$\\
%\cite{Belle:2021avh}\\
$10^{2}\frac{\mathcal{B}(\Xi_c^0\to \Sigma^+ K^-)}{\mathcal{B}(\Xi_c^0\to \Xi^- \pi^+)}$&
$12.3\pm1.2$&&&$12.3\pm1.2$\\
%\cite{Belle:2021avh}\\

$10^{2}\mathcal{B}(\Xi_c^+\to \Xi^0 \pi^+)$&
$1.6\pm0.8$&&&
$1.6\pm0.8$\\

$\alpha(\Lambda_c^+\to \Lambda^0 \pi^+)$&
$-0.84\pm0.09$&&$-0.755\pm0.006$ \cite{Belle:2022uod}&
$-0.76\pm0.01$ ~\cite{Workman:2022ynf,Belle:2022uod}\\

$\alpha(\Lambda_c^+\to \Sigma^0 \pi^+)$&
$-0.73\pm0.18$&&$-0.463\pm0.018$ \cite{Belle:2022uod}&
$-0.47\pm0.03$ ~\cite{Workman:2022ynf,Belle:2022uod}\\

$\alpha(\Lambda_c^+\to p K_S)$&
$0.18\pm0.45$&&&
$0.18\pm0.45$\\

$\alpha(\Lambda_c^+\to \Sigma^+ \pi^0)$&
$-0.55\pm0.11$&&$-0.48\pm0.03$ \cite{Belle:2022bsi}&
$-0.49\pm0.03$ \cite{Workman:2022ynf,Belle:2022bsi}\\

$\alpha(\Xi_c^0\to \Xi^- \pi^+)$&
$-0.64\pm0.05$&&&
$-0.64\pm0.05$\\
%\cite{Belle:2021crz}\\

$\alpha(\Lambda_c^+\to \Sigma^+ \eta)$&
&&$-0.99\pm0.06$ \cite{Belle:2022bsi}&
$-0.99\pm0.06$ \cite{Belle:2022bsi}\\

$\alpha(\Lambda_c^+\to \Sigma^+ \eta')$&
&&$-0.46\pm0.07$ \cite{Belle:2022bsi}&
$-0.46\pm0.07$ \cite{Belle:2022bsi}\\

$\alpha(\Lambda_c^+\to \Lambda^0 K^+)$&
&&$-0.585\pm0.052$ \cite{Belle:2022uod}&$-0.585\pm0.052$ \cite{Belle:2022uod}
\\

$\alpha(\Lambda_c^+\to \Sigma^0 K^+)$&
&&$-0.55\pm0.20$ \cite{Belle:2022uod}&$-0.55\pm0.20$ \cite{Belle:2022uod}\\

$\alpha(\Lambda_c^+\to \Xi^0 K^+)$&
&$0.01\pm0.16$ \cite{BESIII:2023wrw}&&
$0.01\pm0.16$ \cite{BESIII:2023wrw}\\
\hline
\end{tabular}
}
\end{center}
\end{table}

\newpage
%%%%%%%%%%%%%%%%%%%%%%%%%%%%%%%%%%%%%%%%%%%%%%%%%%

%nocite{*}
%\bibliography{reference}% Produces the bibliography via BibTeX.

\end{document}